\documentclass{emulateapj}

\usepackage{graphicx}

\usepackage{amsbsy}

\def\bs{\boldsymbol}
\def\gsim{\;\rlap{\lower 2.5pt
\hbox{$\sim$}}\raise 1.5pt\hbox{$>$}\;}
\def\lsim{\;\rlap{\lower 2.5pt
\hbox{$\sim$}}\raise 1.5pt\hbox{$<$}\;}

\lefthead{Pan et al.}  \righthead{Turbulent Clustering}
\begin{document}
\title{Turbulent Clustering of Protoplanetary Dust and Planetesimal Formation}

\author{Liubin Pan\altaffilmark{1}, Paolo Padoan\altaffilmark{2}, John Scalo\altaffilmark{3}, Alexei G. Kritsuk\altaffilmark{4}, Michael L. Norman\altaffilmark{4}}
\altaffiltext{1}{School of Earth and Space Exploration,  Arizona State University, P.O.  Box 871404, Tempe, AZ, 85287; liubin.pan@asu.edu}
\altaffiltext{2}{ICREA \& ICC, University of Barcelona, Marti i Franqu\`{e}s 1, E-08028 Barcelona, Spain;
ppadoan@icc.ub.edu}
\altaffiltext{3}{Department of Astronomy, University of Texas, Austin, TX 78712}
\altaffiltext{4}{Department of Physics, University of California, San Diego, 
CASS/UCSD 0424, 9500 Gilman Drive, La Jolla, CA 92093-0424}

\begin{abstract}

We study the clustering of inertial particles in turbulent flows and 
discuss its applications to dust particles in protoplanetary disks. 
Using numerical simulations, we compute the radial distribution 
function (RDF), which measures the probability of 
finding particle pairs at given distances, and the probability 
density function of the particle concentration. The clustering 
statistics depend on the Stokes number, $St$, defined as the 
ratio of the particle friction timescale, $\tau_{\rm p} $, to the 
Kolmogorov timescale in the flow. In agreement with previous studies, 
we find that, in the dissipation range, the clustering intensity 
strongly peaks at $St \simeq 1$, and the RDF for $St \sim 1$ 
shows a fast power-law increase toward small scales, 
suggesting that turbulent clustering may considerably 
enhance the particle collision rate. Clustering at inertial-range 
scales is of particular interest to the problem of planetesimal 
formation. At these large scales, the strongest clustering 
is from particles with $\tau_{\rm p}$ in the inertial 
range. Clustering of these particles occurs primarily 
around a scale where the eddy turnover time is 
$\sim \tau_{\rm p}$. We find that particles of different 
sizes tend to cluster at different locations, leading 
to flat RDFs between different particles at small scales. 
In the presence of multiple particle sizes, the overall 
clustering strength decreases as the particle size distribution broadens.  
We discuss particle clustering in two recent models for planetesimal 
formation. We point out that, in the model based on turbulent clustering 
of chondrule-size particles, the probability of finding strong clusters that can seed 
planetesimals may have been significantly overestimated. 
We discuss various clustering mechanisms in  simulations 
of planetesimal formation by gravitational collapse of 
dense clumps of meter-size particles, in particular 
the contribution from turbulent clustering due to the limited 
numerical resolution.
 
\end{abstract}


\section{Introduction}

Dust grains with microscopic to millimeter size are an important component of many 
astrophysical environments, and perhaps most significantly of protoplanetary disks.  Although they contain a small 
mass fraction (approximately 1\% with no gas-grain separation), solid particles affect the gas dynamics 
and emission through various processes such as thermal exchange, surface chemistry, and radiative transfer.  
In protoplanetary disks, their migration, sedimentation, and collisional coalescence and fragmentation set the 
stage for planet formation. Solid particles are dragged by gas motions, which are generally turbulent 
in astrophysical systems. The drag force of the gas turbulence, along with the generic feature that 
the inertial particle trajectories are dissipative, gives these particles a complex 
dynamics consisting of stochastic accelerations and decelerations, resulting 
in motions  that partially reflect features of the velocity field of the gas that 
carries them. 

The effect of turbulence on particle or droplet growth has been studied for over half a century (Arenberg 1939, East \&Marshall 1954), 
and remains a challenging problem today in many research fields, particularly in the study of turbulent atmospheres. It is relevant 
to cloud formation, rain initiation (see Shaw (2003) for a general review),  and the general microphysics (Pruppacher and Klett 1998) 
of the atmospheres of planets and moons ( e.g. Barth and Rafkin 2007, McGouldrick and Toon 2008), and of cool stars and brown 
dwarfs (Helling and Woitke 2006, Helling et al.\ 2008, Marley, Didier, and Goldblatt 2010, Freytag et al.\ 2010). 

For disks, an important effect is that turbulent motions can induce random relative velocities between inertial particles 
that are much larger than Brownian velocities, increasing the particle collision rates, and hence growth rates, but also 
leading to destructive collisions if the relative particle speed exceeds a threshold believed to be of order a few 
cm/sec (see Blum and Wurm (2008) for a review; Guttler et al.\ (2010) for an update). In the present 
paper we focus on another aspect of the coupling of turbulence with solid particles in disks: 
Turbulent clustering. Because the inertia of particles prevents a perfect coupling with the flow, 
dissipative trajectories forced by turbulence can cause the formation of dense clusters of 
particles, even if the flow is incompressible. The process is sometimes referred to as ``preferential concentration'' 
(Fessler et al.\ 1994) in atmospheric and engineering applications. 

The ability of incompressible turbulence to generate clusters of small particles was suggested in a 
seminal paper by Maxey (1987), and has been confirmed both numerically (Squires \& Eaton 1991;  Wang \& Maxey 1993)
and experimentally (Fessler et al.\ 1994, Uhlig et al.\ 1998, Kostinski \& Shaw 2001, Aliseda et al.\ 2002, Pinsky \& Khain 2003, 
Wood et al.\ 2005, Lehmann et al.\ 2007). 
The basic features of turbulent clustering were established in a number of theoretical studies 
(Elperin et al.\ 1998a, Elperin et al.\ 1998b, Balkovsky et al.\ 2001, Zaichik et al.\ 2003a, Zaichik et al.\ 2003b, Elperin et al.\ 2002) 
and low-resolution simulations  (Sundaram \& Collins97, Zhou et al.\ 1998, Reade \& Collins 2000a, Reade \& Collins 2000b, Wang et al.\ 2000). 
Most of these studies focused on clustering at the dissipation-range scales. In this scale range, the clustering intensity was found to peak for 
particles with Stokes number (the ratio of particle friction time to the Kolmogorov timescale) close to unity, 
and the clustering amplitude was shown to increase towards smaller scales as a power law. Higher-resolution 
turbulence simulations (Hogan et al.\ 1999, Hogan\& Cuzzi 2001, Collins \& Keswani 2004, Falkovich \& Pumir 2004, Bec et al.\ 2006a, Cencini et al.\ 2006)
have confirmed these basic results, but still differ concerning the scaling of the clustering amplitude with the Stokes 
and Reynolds numbers. 

The process of turbulent clustering has been proposed as a possible solution to the problem of raindrop formation 
in atmospheric clouds (Jameson\& Kostinski 2000, Falkovich et al.\ 2002, Vaillancourt et al.\ 2002), due to its effects on the collision 
rate of droplets. As in the case of droplet formation, the collision rate between dust grains in astrophysical 
systems may be enhanced by turbulent clustering.  A major goal of this paper is a general introduction of 
the phenomenon of turbulent clustering to the astronomy community, presenting a detailed physical discussion and numerical results.  
We also discuss the application of our simulation results to models of planetesimal formation 
in protoplanetary disks.
 
Planetesimals are kilometer-size objects believed to be the necessary 
precursors to the formation of fully-fledged rocky planets. The classic theory assumes that 
planetesimals form by gravitational instability, as the dust 
particles vertically settle to a dense thin layer at the midplane (Safronov 1969, Goldreich and Ward 1973). 
However, even without preexisting turbulence, 
size-differentiated sedimentation of the particles results in 
vertical shear that can lead to Kelvin-Helmholtz instabilities 
as suggested by Weidenschilling (1980) (see Barranco 2009 for a recent detailed study). 
The resulting turbulent mixing prevents the settling to a thin dust layer, and the dust 
density needed for the gravitational instability to occur may be difficult to achieve 
(e.g. Youdin and Shu 2002, Chiang 2008). Another possibility is that planetesimals form
by the collisional growth of dust particles. Early work on collisional 
growth of planetesimals and planets was reviewed by Lissauer (1993). 
The most serious problem for planetesimal formation in a turbulent disk 
continues to be that both theoretical (e.g. Ormel et al.\ 2007, Brauer et al.\ 2008) and 
experimental (see Blum and Wurm 2008 for a thorough review) studies 
indicate that particle growth is stalled in the cm-m size range, 
a conundrum usually referred to as the meter-size problem. 
Fast radial migration of cm-m particles could be alleviated with a 
modest enhancement of the dust-to-gas ratio, but these 
particles acquire such large velocities that collisional fragmentation appears 
inevitable (see Brauer et al.\ 2008).  A recent summary of work on planetesimal 
growth is presented by Chiang and Youdin (2010), who emphasize the possibility 
that drag instabilities can concentrate particles and initiate gravitational 
instability of particle clusters (Goodman and Pindor 2000, 
Youdin and Goodman 2005, Johansen et al.\ 2007). 

One response to these problems is to use them to argue that 
turbulence must not exist. Another is to accept one of several 
mechanisms (see Chiang and Youdin 2010) suggested to avoid 
the meter-size problem. Some of these mechanisms are based 
on the formation of dense particle clumps by the clustering of particles by the disk 
turbulence (Cuzzi et al.\ 2008), or by the streaming instability and other clustering effects 
(Johansen et al.\ 2007, 2009a, 2011).  The point of view of the present paper is 
to take a critical look at the aspects of the models that rely on clustering of small particles 
as a part of planetesimal formation, using a new high-resolution turbulence simulation, 
along with a set of approximate guidelines to the behavior we find.  
 
The paper is organized as follows. \S2 is a general introduction to the physics 
of turbulent clustering.  In \S 3 we describe our numerical  simulations.  We 
present results on the clustering statistics of identical particles in \S 4. 
In this section, we also discuss the Reynolds number dependence and 
possible effects of the back reaction, largely based on a review 
of numerical results from the literature. The clustering statistics of particles 
of different sizes are presented in \S5. We apply our understanding of turbulent 
clustering to the problem of planetesimal formation in \S6, with specific discussions 
of the models by Cuzzi et al.\ (2008) and Johansen et al.\ (2007). 
Our conclusions are summarized in \S 7.

\section{Inertial Particle Clustering in Turbulent Flows}

In order to guide the interpretation of the numerical results, we present here a 
brief introduction to the problem of particle clustering. We show how simple physical
arguments allow us to make rough predictions about the Stokes number 
dependence of turbulent clustering that will be computed later from our 
numerical simulation. 

The velocity, $\bs{v}(t)$, of an inertial particle 
suspended in a turbulent velocity field,  
$\bs{u} (\bs{x}, t)$, is given by the equation, 
\begin{equation}
\frac {d \bs{v} } {dt} = \frac { \bs{u} (\bs{x}_{\rm p}(t), t) - \bs{v}} 
{\tau_{\rm p}}            
\end{equation} 
where $\bs{u}(\bs{x}_{\rm p}(t), t)$ is the flow velocity 
along the particle trajectory, $\bs{x}_{\rm p}(t)$, and 
the friction timescale, $\tau_{\rm p}$, represents 
the particle inertia and is essentially the time needed 
for the particle velocity to relax toward the flow 
velocity through the friction force. 

The estimate of the friction timescale depends on 
the particle size, $a_{\rm p}$, relative to the mean 
free path of the gas molecules, $\lambda_{\rm g}$, 
in the flow (see, e.g., Weidenschilling 1977; Cuzzi et al.\ 1993). If 
$a_{\rm p} \ll \lambda_{\rm g}$, the particle-flow 
friction is in the Epstein Regime where the drag 
force is controlled by collisions between the particle 
and the flow molecules. The friction time is calculated by,  
\begin{equation}
\tau_{\rm p}=\left( \frac{\rho_{\rm d}}{\rho_{\rm g}}\right)
\left(\frac{a_{\rm p}}{C_{\rm s}} \right)
\end{equation} 
where  $C_{\rm s}$ is the gas thermal velocity, 
$\rho_{\rm g}$ is the density of the flow and 
$\rho_{\rm d}$ is the density of the particle 
material. For compact dust grains, $\rho_{\rm d} \sim 1$~g~cm$^{-3}$.   
The gas mean free path is estimated by $1 (\rho_{\rm g}/10^{-9}$~g~cm$^{-3}$)$^{-1}$ cm, 
assuming the cross section of hydrogen molecules is $\sim 10^{-15}$ cm$^2$.  
Therefore, the friction between dust particles and the flow 
is in the Epstein regime for particle size up to $\sim 1 (\rho_{\rm g}/10^{-9}$~g~cm$^{-3}$)$^{-1}$ cm.
Due to the density dependence, this critical size 
varies with the radial locations in the disk and 
depends on the disk parameters.  

On the other hand, for particles with $ a_{\rm p} \gg \lambda_{\rm g}$, 
the friction force is determined by the flow 
around the particle surface. If the flow around the 
particle is laminar,  the friction timescale is 
given by the Stokes law,
\begin{equation}
\tau_{\rm p}=\frac{2}{9} \left( \frac{\rho_{\rm d}}{\rho_{\rm g}} \right) \left(\frac{a_{\rm p}^2}{\nu}\right) 
\end{equation} 
where $\nu$ is the kinematic viscosity of the carrier flow.  

The Stokes number, $St$, defined as the ratio 
of the friction timescale to the Kolmogorov timescale, 
$\tau_{\eta}$, i.e., $St \equiv \tau_{\rm p}/\tau_\eta$,  
is commonly used to characterize the particle inertia. 
The Kolmogorov timescale is essentially the turnover 
time of the smallest eddies and is thus the smallest 
timescale in a turbulent flow. It is defined as $\tau_\eta = (\nu/\bar{\epsilon})^{1/2}$ 
where $\bar{\epsilon}$ is the average energy dissipation rate.  
In incompressible turbulence, we have 
$\bar{\epsilon} = \nu \langle {\bf \omega}^2 \rangle$  
with ${\bf \omega}$ being vorticity, and thus 
$\tau_{\eta}$ can be calculated as $\tau_{\eta}= \langle \omega^2 \rangle^{-1/2}$. 
It can also be roughly estimated from the large-scale 
properties of the flow by $\tau_{\eta}  \simeq (L/U)Re^{-1/2}$ 
where $L$, $U$ and  $Re$ are, respectively, 
the outer length scale, the rms flow velocity and the 
Reynolds number. A crucial length scale in 
the clustering statistics of inertial particles is 
the Kolmogorov dissipation scale, $\eta$, 
which is given by $\eta \equiv (\nu^3/ \bar{\epsilon})^{1/4} \simeq L Re^{-3/4}$. 
Numerical values for these quantities applicable to disks are given in \S6.1.

The spatial clustering of inertial particles in 
turbulent flows has different behaviors 
for $St < 1$ and $St >1$.  We discuss 
the two Stokes number ranges separately.

\subsection{Particles with $St<1$}

The trajectories of small particles with $St \ll 1$ 
deviate from those of the fluid elements only 
slightly, and the particle phase can be 
approximately described as a fluid. 
The velocity field, $v_i(\bs{x}, t)$, 
of the particle flow can be estimated from eq.\ (1). 
Assuming that the particle acceleration, 
$\frac{dv_i}{dt}$, can be approximated by the local flow 
acceleration, $\frac{du_i}{dt}$, we have $v_i(\bs{x}, t) 
\simeq u_i(\bs{x}, t) - \tau_{\rm p} \frac{du_i}{dt}(\bs{x}, t)$. 
The assumption is justified for  $St \ll 1$ 
particles because the friction timescale 
$\tau_{\rm p}$ is smaller than $\tau_{\eta}$, 
the smallest timescale in the flow.  
The approximation is essentially the Taylor 
expansion of eq.\ (1) to the first order of $St$. 

With this approximation, one can estimate 
the divergence, $\partial_i v_i$, of 
the particle velocity field. If the carrier 
flow is incompressible, we have, 
 \begin{equation}
\partial_i v_i = -\tau_{\rm p} \partial_i u_j \partial_j u_i,  
\end{equation} 
where we used $\frac{du_i}{dt} = \frac{\partial u_i}
{\partial t} + u_j \partial_j u_i$ and $\partial_i u_i=0$.
Eq.\ (4) suggests that the particle 
flow has a finite compressibility 
even though the carrier flow is incompressible, 
and this would lead to spatial clustering of 
the particles. Intuitively, the physical origin for 
clustering is that the particles' inertia causes 
them to lag behind or lead in front of the flow elements 
when the flow experiences an acceleration or deceleration.    

The amplitude of the particle velocity divergence 
depends on the flow velocity gradient. 
On average, the velocity gradient in a turbulent flow 
is $\sim (\bar{\epsilon}/\nu)^{1/2} = 1/\tau_\eta $ 
(e.g., Monin and Yaglom 1975), thus we 
have an estimate that $\partial_i v_i \simeq St/\tau_\eta$. 
In the limit of small Stokes numbers, the 
divergence increases with increasing $St$, 
and thus the degree of clustering is expected to 
increase with  $St$.

The particle velocity divergence can be 
rewritten as $ \tau_{\rm p} (\omega^2/2-s_{ij}s_{ij})$ 
where $s_{ij} =(\partial_i u_j + \partial_j u_i)/2$ 
is the strain tensor (Maxey 1987). 
This suggests that vorticity tends to expel particles, 
while the strain would collect particles. 
Therefore dense particle clusters are expected 
to be found in the strain-dominated regions with 
low vorticity.  This effect is illustrated in Appendix A where 
we use Burgers vortex tubes as a model for the  
small-scale structures in turbulent flows.  The effect of
vortices as centrifuges for inertial particles  
was first recognized by Maxey 
(1987), and has been subsequently studied 
in details with both numerical simulations (e.g., 
Wang and Maxey 1993) and experiments 
(e.g., Fessler et al.\ 1994). 

Eq.\ (4) can also be written as 
$\partial_i v_i = \tau_{\rm p} \partial_i^2 P/\rho_{\rm g}$.
This means that the particle flow divergence is negative 
at local pressure maxima where $\partial_i^2 P/\rho_{\rm g}<0$. 
Therefore, particle clustering in turbulent flows is 
sometimes interpreted as collection of particles 
at local pressure maxima.

The velocity gradient field in a turbulent 
flow has a correlation length scale of 
$\eta$, and thus the divergence of the particle 
flow is decorrelated at scales larger than $\eta$. 
Therefore the probability for the existence 
of coherent particle compressions or expansions 
at scales significantly larger than $\eta$ would 
be rare, suggesting that, at $St \lsim 1$, particle 
clustering would primarily occur below the 
Kolmogorov length scale. However, this 
does not mean that the particle clusters 
appear as spheres of size $\sim \eta$. Instead, 
they are found to be in the form of filaments or sheets 
of thickness $\sim \eta$. 

Particle clusters are subject to disruption by the 
stretching of the carrier flow, which tends to disperse the 
clusters. The balance between the disruption and the 
compressibility in the particle flow determines the clustering intensity. 
At smaller scales, it takes longer time for stretching to disperse 
particle clusters to scales larger than $\eta$ 
where essentially no coherent compressions 
or expansions exist. Therefore a higher level 
of clustering is expected at smaller scales 
because clusters at these scales can experience 
coherent compressions for longer time 
(Falkovich and Pumir 2004).    

It is interesting to note that the quadratic 
dependence of the particle flow divergence 
on the velocity gradients is similar to that of 
the energy dissipation rate $\epsilon({\bs x}, t)
=\frac{1}{2} \nu (\partial_i u_j + \partial_j u_i)^2 $. 
Therefore, like the dissipation rate, $\partial_i v_i$ 
would also display spatial fluctuations, 
which may give rise to a broad probability 
density function (PDF) for the particle 
concentration. Also, it is known that the PDF of the energy 
dissipation rate broadens with increasing 
Reynolds number (Frisch 1995). 
A similar Reynolds number dependence is 
likely to exist for the concentration 
PDF of particles with $St \lsim 1$. 

\subsection{Particles with $St>1$}

With increasing inertia, the particle trajectories 
deviate more from those of the flow elements. 
A large particle has a long memory, and its 
current velocity has significant contribution 
from the memory of the flow velocity in the 
past. Therefore, the particle velocity cannot be 
simply estimated by the local carrier flow. 
The approximation, eq.\ (4), for the 
particle flow divergence breaks down 
for $St$ larger than 1. 

In fact, nearby large particles with $St \gg 1$ 
do not move coherently, and at small scales 
the particle phase can no longer be viewed 
as a fluid. Intuitively, due to their large inertia,
two large particles can keep a significant 
relative speed when approaching each other. 
Therefore the relative particle motions 
at small scales appear to be random.
Bec et al.\ (2010) found that, for $St >1$, 
the velocity difference, $\delta v(St, l)$, 
of two particles at a separation $l$ 
is constant at small values of $l$, indicating 
that their relative motions are similar to the 
thermal motions of molecules in kinetic theory. 
Thus a fluid description for these particles 
would not be sufficient. The physical reason 
for a constant $\delta v (St, l)$ at small $l$ (for a given $St$) 
is that the relative velocity between nearby particles 
is dominated by their memory of the flow velocity 
difference they ``saw" within a friction 
timescale in the past (Pan and Padoan 2010). 
 
We consider the response behavior of $St >1$ 
particles to turbulent eddies of different sizes, 
which provides physical insights to 
the clustering properties of these particles. A length scale 
of particular interest is the size of turbulent 
eddies whose turnover timescale is equal to the 
particle friction timescale, $\tau_{\rm p}$. 
If $\tau_{\rm p}$ corresponds to an inertial-range 
timescale of the carrier flow, we have 
$l_{\tau_{\rm p}} \simeq \bar{\epsilon}^{1/2} \tau_{\rm p}^{3/2}$
(or equivalently $\simeq St^{3/2} \eta$) using the Kolmogorov scaling. 

Particles can efficiently respond to  eddies much 
larger than $l_{\tau_{\rm p}}$.  At these scales, 
the particle motions are well coupled to the flow 
elements, and the particle velocity difference, 
$\delta v (St, l)$, essentially follows the flow 
velocity difference, $\delta u (l)$ (see Bec et al.\ 2010). 
Therefore, no strong particle clustering is expected 
at these large scales. Eddies much smaller 
than $l_{\tau_{\rm p}}$ do not efficiently 
affect the relative particle motions because the 
particle response time, $\tau_{\rm p}$, is 
much longer than the eddy turnover time. 
Thus, at scales below $l_{\tau_{\rm p}}$, the 
flow and the particle motions are decoupled, 
and the relative velocity between two particles 
is determined by their memory of the flow 
velocity difference at scales around $l_{\tau_{\rm p}}$, where 
the particle motions are partially coupled to the carrier flow.  As discussed above, 
particles show random relative motions at these small scales, and thus no clustering 
would be found at $l \ll l_{\tau_{\rm p}}$ either.  This means that 
significant clustering could occur only around the scale 
$\sim l_{\tau_{\rm p}}$. This physical picture also suggests that 
the particle phase has an effective mean free path of $l_{\tau_{\rm p}}$. A fluid description for 
particles may be valid at scales above $l_{\tau_{\rm p}}$. 
  
For particles with $\tau_{\rm p}$ larger than the turnover time, $T_{\rm L}$, at the outer 
scale of the flow, all eddies evolve at a timescale smaller than $\tau_{\rm p}$, and $l_{\tau_{\rm p}}$ 
cannot be defined. Such particles do not closely follow the flow velocity at 
any scale.  Motions of these particles are expected to be random at all scales, and the 
spatial distribution would be essentially homogeneous. We focus on inertial-range
particles with $\tau_\eta \ll \tau_{\rm p} \ll T_{\rm L}$ in our discussions.

The clustering intensity for inertial-range particles is 
expected to decrease with increasing $St$.  As discussed 
above, these particles cluster primarily at the length scale, 
$l_{\tau_{\rm p}}$, which increases with $St$. 
Therefore, clusters of larger particles are spatially 
more spread out, and, since no strong fluctuations 
exist  below $l_{\tau_{\rm p}}$,  the concentration level 
within the clusters would decrease with increasing $St$. 
In other words, smaller particles can form thinner 
clusters with higher density contrast and hence exhibit stronger clustering.  
The decrease of the clustering intensity with $St$ is
illustrated by an intuitive example in Appendix A. The 
example shows that  larger particles (with $St>1$)  
form clusters of larger sizes, and the particle 
concentration in the clusters becomes smaller 
with increasing $St$.    

We estimate the compressibility in the particle collective 
motions around the scale $l_{\tau_{\rm p}}$, 
which is used in Appendix B for the derivation 
of the Brownian scale.  Here the scale $l_{\tau_{\rm p}}$ 
is of special interest 
because the maximum flow velocity gradient 
that the particles can efficiently ``feel" is that at $l_{\tau_{\rm p}}$. 
The gradient is approximately $\delta u(l_{\tau_{\rm p}})/l_{\tau_{\rm p}}$, 
which is $\propto \bar{\epsilon}^{1/3} l_{\tau_{\rm p}}^{-2/3}$ 
using the Kolmogorov scaling. The 
gradient decreases as $(\tau_{\rm p})^{-1}$ with $\tau_{\rm p}$, 
which also suggests weaker clustering for larger particles. 
The divergence of the particle motions around the scale 
$l_{\tau_{\rm p}}$ is calculated by the same method 
(eq.\ 4) as for the $St<1$ particles. This is justified because 
the friction time is smaller than the turnover time of  eddies 
larger than $l_{\tau_{\rm p}}$. Inserting $\delta u(l_{\tau_{\rm p}})/l_{\tau_{\rm p}}$ 
for the velocity gradients in eq.\ (4) shows that the effective 
divergence is $\sim \tau_{\rm p}^{-1} = (St \tau_\eta)^{-1}$. 
Therefore, for $St\gsim 1$, the particle collective motions 
are less compressible as $St$ increases. 

We note that, unlike particles with $St <1$, 
the effective divergence estimated above for 
$St>1$ only depends on the particle friction time, 
but not on the flow properties in the inertial range. 
This is because clustering of these particles 
occurs at scales ``selected" by the particle 
timescale.  At the selected length scale, 
the turnover timescale is around $\tau_{\rm p}$, 
and the flow velocity gradient is $\sim \tau_{\rm p}^{-1}$. 
It is thus not surprising that the effective divergence
is determined solely by the friction timescale. The 
possibility of clustering of large particles at an 
inertial-range scale $\sim l_{\tau_{\rm p}}$ 
has also been discussed in earlier studies 
(e.g., Eaton and Fessler 1994, Boffetta et al.\ 2004, 
Bec et al.\  2007).  
\\

In summary,  inertial particles suspended in a 
turbulent flow are expected to show inhomogeneous 
spatial distribution even if the carrier flow is 
incompressible. Inertial particles tend to be expelled from 
vortices and accumulate in high-stain regions. For small particles 
with $St \lsim 1$, clustering occurs primarily at 
scales below the Kolmogorov scale $\eta$, 
and the degree of clustering increases with 
increasing $St$. Large particles with $1 \lsim St \lsim T_{\rm L}/\tau_{\eta} $ 
cluster around a scale, $l_{\tau_{\rm p}}$, which 
increases with $St$ as $\simeq St^{3/2} \eta$. 
The clustering intensity decreases with $St$ for 
$St \gsim 1$.  Overall, the clustering intensity 
is expected to peak at  $St \sim 1$.

\subsection{Clustering of particles of different sizes} 

The discussion above is for particles 
of the same size, an idealized situation 
usually referred to as the monodisperse 
case. In realistic environments, the 
particle size is likely to have significant 
variations either due to an initial size 
distribution (from the formation process 
of the particles) or as a result of collisional 
coagulation or fragmentation. Therefore 
it is necessary to consider the clustering 
statistics for particle of different sizes.

Numerical simulations by Zhou et al.\ (2001) 
showed that particles with different sizes  
tend to cluster at different locations in 
the flow (see also Reade and Collins 2000b). 
This is also clearly illustrated by our 
example in Appendix A. A consequence of 
this effect is that the probability of finding 
nearby particles of a different size is smaller 
than that of finding identical particles, 
given equal number densities of the two particles. 
This has interesting effects on the collision 
kernel for particle coagulation models 
(Reade and Collins 2000b). 
It also has important implications on the overall 
spatial distribution of particle density/concentration 
when the particles have an extended size range. 
A detailed analysis of the clustering statistics 
for particles of different sizes will be given in \S 5.

\section{Numerical Simulations}

With the rough but physically-motivated arguments of \S2 in 
hand, we now present and interpret the results of our numerical 
simulation. The simulation was carried out in a periodic box 
with $512^3$ grid points. The hydrodynamic equations with 
an isothermal equation of state were solved by the Enzo 
code (O Shea et al.\ 2004 and references therein), which 
uses a direct Eulerian formulation of the Piecewise 
Parabolic Method (PPM) (Colella and Woodward 1984). 
To drive the turbulent flow and maintain the kinetic 
energy at the desired level, we apply a large-scale 
solenoidal force with a fixed spatial pattern and a 
constant power in the range of wave numbers $1\le k \le 2$. 
The amplitude of the driving force is chosen such that 
the rms Mach number, $M_{\rm s}$, in the flow is $\simeq 1$
(the simulation setup is the same as Kritsuk 
et al.\ (2007), except for the lower Mach number 
and the solenoidal forcing adopted here). Unlike previous simulations 
devoted to exploring particle clustering in incompressible 
turbulence (e.g. Sundaram \& Collins 1997; Reade \& Collins 2000a; Hogan \& Cuzzi 2001; 
Collins \& Keswani 2004; Falkovich \& Pumir 2004; 
Cencini et al.\ 2006), our simulated flow is compressible.

We chose to study turbulent clustering with a 
compressible flow because we aimed at 
exploring dust grain dynamics in various environments 
including highly compressible interstellar clouds. In the 
current work, we will focus on the application in 
proptoplantary disks where the turbulence is essentially 
incompressible. We expect from the following considerations that 
the clustering statistics in our simulated flow would be close to 
that in incompressible turbulence.  First, at Mach number 
close to unity, the density fluctuations are weak, with 
the rms amplitude $\langle \delta \rho_{\rm g}^2 \rangle^{1/2}/\bar{\rho}_{\rm g}$ 
at the level of $\sim$10 percent. Second,  the velocity 
structures in a transonic flow are very close to those in 
incompressible flows (Porter et al.\ 2002; Padoan et al.\ 2004; 
Pan and Scannapieco 2011). 
In \S 4.1, we find that the clustering properties in our 
transonic flow are indeed in good agreement with 
the results from direct numerical simulations (DNS) for 
incompressible flows by Collins \& Keswani (2004).  This agreement  
validates the application of our results to protoplanetary disks.

One important quantity in our statistical analysis is the 
Kolmogorov length scale. This length scale is difficult to 
evaluate because our PPM simulations  do not 
explicitly include the viscous term and the kinetic energy dissipation 
is through numerical diffusion. We compute $\eta$ using 
two methods. In the first method, we start with an estimate 
of the effective viscosity, $\nu_{\rm eff}$. We calculate 
$\nu_{\rm eff}$ from the equation 
$\bar{\epsilon} = \nu_{\rm eff} \langle \omega_i^2 \rangle$, 
because solenoidal modes dominate the kinetic 
energy dissipation even in a transonic flow 
(Pan and Scannapieco 2010). The energy 
dissipation rate, $\bar{\epsilon}$, can be derived 
either from Kolmogorov's 4/5 law (which also applies 
also to transonic flows;  see Pan and Scannapieco 2010 
and also Benzi et al.\ 2008),  
or from the relation, $\bar{\epsilon}= D u'^3/L_1$, established 
by DNS, where $u'$ and $L_1$ are the 1D velocity 
dispersion and the integral length scale, 
and the coefficient $D \simeq 0.4$ (Ishihara et al.\ 2009). 
The dissipation rate values derived from the two approaches 
are consistent with each other.  The effective viscosity 
$\nu_{\rm eff}$ is then calculated from $\bar{\epsilon}$ and $\langle w^2 \rangle$.  
With $\nu_{\rm eff}$, we find the effective Taylor Reynolds number 
in our simulated flow is $Re_{\lambda} =250$. We calculate the
Kolmogorov length scale from $\eta=(\nu_{\rm eff}^3/\bar{\epsilon})^{1/4}$, 
which turns out to be 1/2 the resolution scale (Benzi et al.\ 2008).
The Kolmogorov timescale is computed  by $\langle w^2 \rangle^{-1/2}$.

\begin{figure}
\centerline{
\includegraphics[width=1.0\columnwidth]{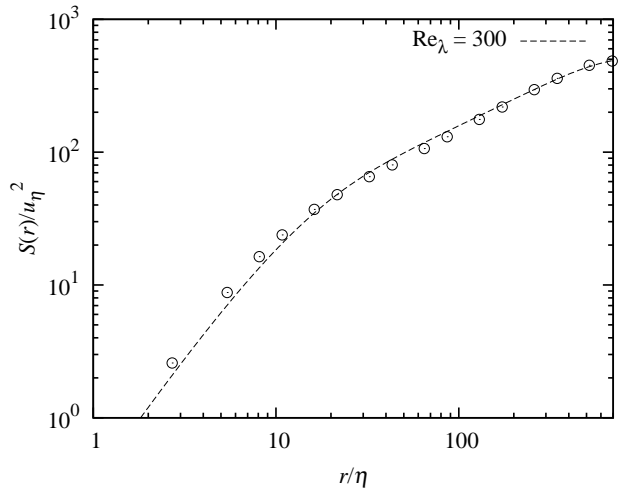}
}
\caption{Second order velocity structure function in our simulated flow (data points).
The dashed line is the structure function for an incompressible flow with $Re_\lambda =300$, 
obtained from a bridging formula that connects the established scaling behaviors in 
different scale ranges (see text).} 
\label{structure}
\end{figure}

In the second method, we estimate $\eta$ by comparing the 2nd 
order velocity structure function, $S(r) = \langle (v_i(\bs{x}+\bs{r},t) - v_i(\bs{x}, t) )^2\rangle$, 
in our flow to that established for incompressible turbulence 
from theory, experiments and simulations. We adjust the 
Kolmogorov scale (or equivalently the effective viscosity) in our flow 
to obtain a best fit. Our result is shown Fig.\ \ref{structure}, where the 
length scale and the structure function are normalized to the Kolmogorov 
scale, $\eta$, and velocity, $u_\eta$, respectively.  The data points 
represent the structure function measured in our simulated flow. The Kolmogorov 
scale is set to be 0.4 times the computation cell size.  With this 
value for $\eta$, we estimated the effective viscosity and 
the Taylor Reynolds number. The latter is $\sim 300$. 
The dashed line is the expected structure function in an incompressible 
turbulent flow with $Re_{\lambda} = 300$. It is obtained from a bridging formula 
given in Zaichik et al.\ (2006), which connects the established scaling 
behaviors of the structure function in different scale ranges.  
Clearly, the data points are in good agreement with the dashed line. 
This agreement suggests that our simulations can be safely 
used for the study of turbulent clustering in weakly 
compressible turbulence such as that in protoplanetary disks. The best-fit value for 
the Kolmogorov scale, 0.4 cell size, is close to that derived from first 
method, suggesting our estimate of $\eta$ is reliable. 
Throughout the paper, we set $\eta$ to be 0.4 cell size  and assume 
the Taylor Reynolds number is 300.

A strength of the PPM method is that it yields 
a quite broad inertial range already at the resolution of $512^3$.   
A clear Kolmogorov scaling is seen in Fig.\ \ref{structure} 
at scales from $\simeq 30\eta$ to $\simeq 300 \eta$ in 
the velocity structure function. To our knowledge, 
turbulent clustering has not been studied in simulations 
that have a clear inertial range. The inertial-range velocity 
scaling was used in our physical discussion in \S2.2.
Our numerical results 
show that the clustering behaviors are different at 
scales below and above the Kolmogorov scale $\eta$. 
We will refer to the scales below $\eta$ as the dissipation range, 
and loosely call the scale range $l > \eta$ the inertial 
range, although the latter usually refers to the scales showing a 
Komolgorov scaling.   

Because the Kolmogorov scale is below the resolution scale, 
one may be concerned with the reliability and accuracy of the 
measured statistics around or below $\eta$. Fortunately,  
we find that the velocity field at the unsolved scales 
may be reliably approximated by interpolation. This is because the velocity structure 
function is already smooth at the resolution scale, as seen 
from the $r^2$ scaling at the smallest scales in Fig.\ \ref{structure}. 
This scaling means that the velocity difference is linear 
with $r$, and a linear interpolation (see below) may 
sufficiently reflect the subgrid velocity statistics. Therefore, 
our simulation can provide good clustering statistics at scales 
around or below $\eta$. This is again supported by the 
agreement of our results  with those from DNS 
simulations (see \S 4.1).   

\begin{figure*}
\centerline{
\includegraphics[width=1.0\columnwidth]{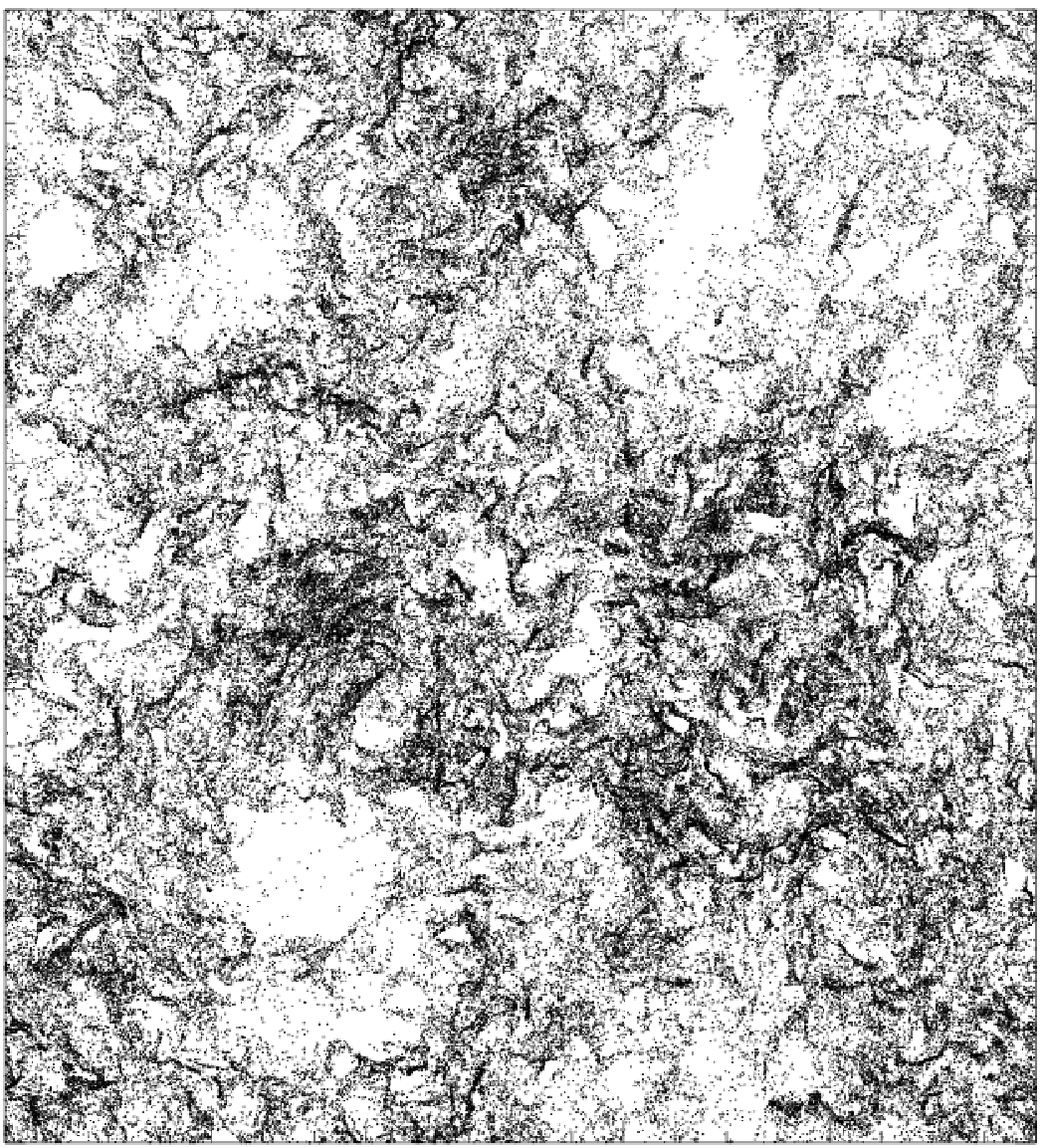}
\includegraphics[width=1.0\columnwidth]{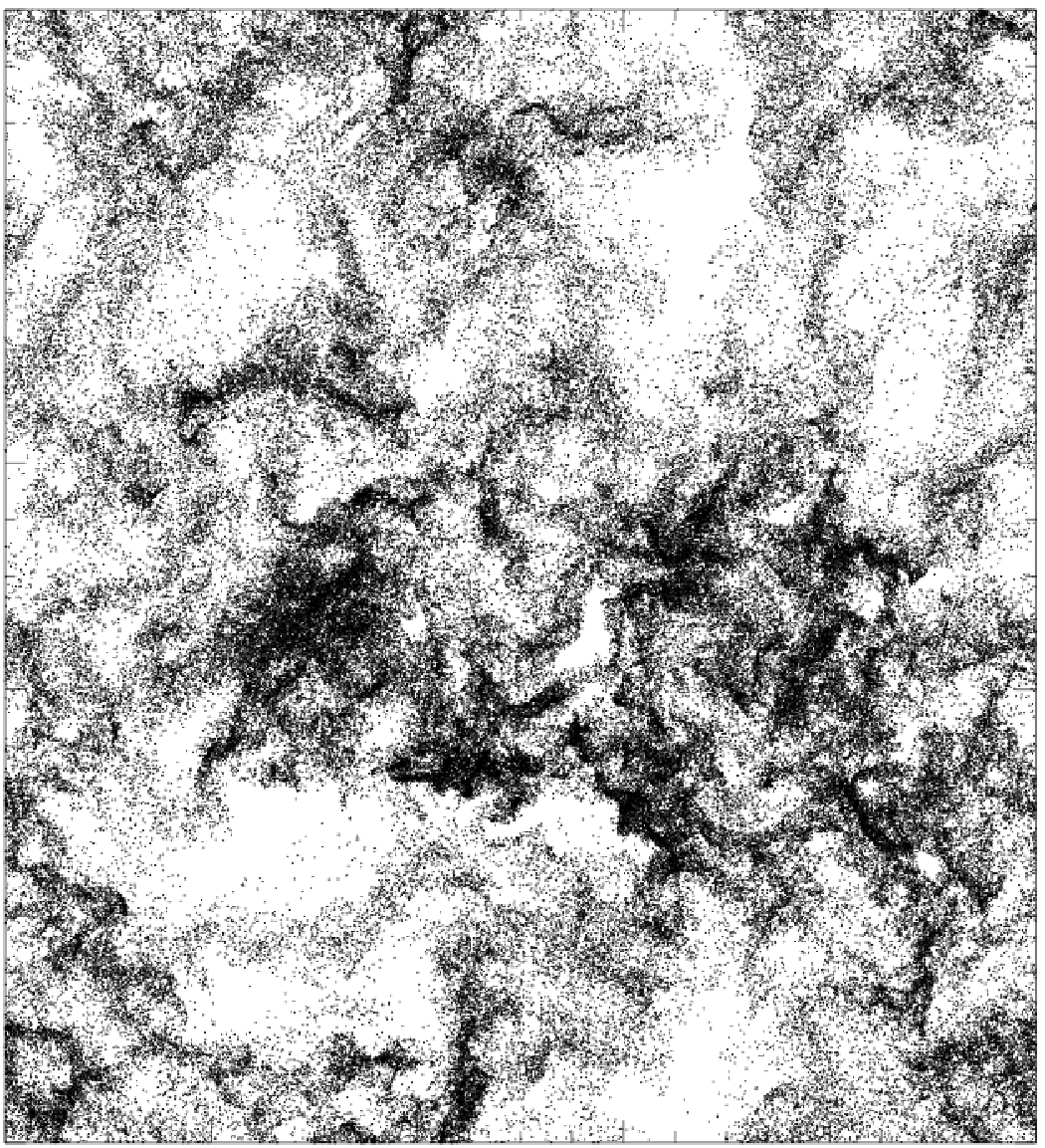}
}
\caption[]{Positions of all the particles with $St=1.2$ (left panel) 
and $St=4.9$ (right panel) within a thin slice of 
thickness equal to 2\% of the computational box. Density fluctuations 
are much stronger for particles with $St=1.2$ than for those 
with $St=4.9$, but this cannot be fully appreciated from the images, 
due to the overlap of particle positions. Notice the very small scale 
structures present in the spatial distribution
of the $St=1.2$ particles. Dense particle filaments and large 
voids can be seen in both panels, 
with sizes approaching the integral length of the flow, 
estimated to be approximately 0.2 times 
the computational box size. The estimated size of the 
dissipation scale, $\eta$, is approximately 
$10^{-3}$ times the box size, as discussed in \S3.}
\label{image}
\end{figure*}     
   
We chose 16 different values for the particle size, and for 
each size we evolved 8.4  million particles in the 
simulated flow. The average particle density for each size is one per 
16 computation cells. Due to the slight density fluctuations in 
our transonic flow, the friction timescale for a particle of a 
given size is not constant along its trajectory. The friction time scales 
with the gas density, $\rho_{\rm g}$, as $\tau_{\rm p} \propto \rho_{\rm g}^{-1}$ 
in the Epstein regime (the regime of primary interest in our 
astrophysical applications), and we calculated the local 
values of $\tau_{\rm p}$ using this scaling at each integration 
step for the particle trajectory. A linear interpolation is used to obtain 
the flow velocity and density at the particle positions inside 
the computation cells. A higher-order interpolation scheme may 
be needed for more accurate measurements of the clustering 
statistics below the resolution scale (e.g., Yeung and Pope 1998).

Like the friction timescale, the Stokes number has 
weak spatial variations. For each particle size, we 
define an average Stokes number using the average 
friction timescale based on the mean flow density. 
The 16 particle sizes cover a Stokes number range 
from 0.08 to 3000. Our statistical analysis will focus on 11 
relatively small particle sizes with $St$ in the 
range $[0.08, 43]$, as the larger particles 
do not show significant clustering. Furthermore, 
the largest particles have a long relaxation time, and their 
statistics may not have saturated at the end of our simulation run.

We neglect particle collisions in our simulations. 
This is a good approximation if the volume filling 
factor, $\Phi_{\rm v}$, is much smaller than 1, which 
is the case for dust particles in astrophysical 
environments. The volume filling factor is 
defined as $\Phi_{\rm v}=\bar{n}_{\rm p} a_{\rm p}^3$, where $\bar{n}_{\rm p}$ 
is the average particle number density.  

The back reaction of the particles on the carrier 
flow is also neglected. The importance of the back 
reaction is measured by the mass loading factor, 
$\Phi_{m} = (\rho_{\rm p}/\rho_{\rm g}) \Phi_{\rm v}$ 
(the ratio of the bulk particle mass density to 
the flow density). On average, $\Phi_{\rm m}$ is 
small, $\sim 0.01$, for dust particles in the 
astrophysical environments with metallicity close to the solar value. 
However, the local mass loading factor could be 
significant in clusters with particle concentration 
much larger than the average. The effect  of mass 
loading should be considered in such clusters. 
We will discuss this effect in more details in \S 4.5.

\begin{figure*}
\centerline{
\includegraphics[width=1.0\columnwidth]{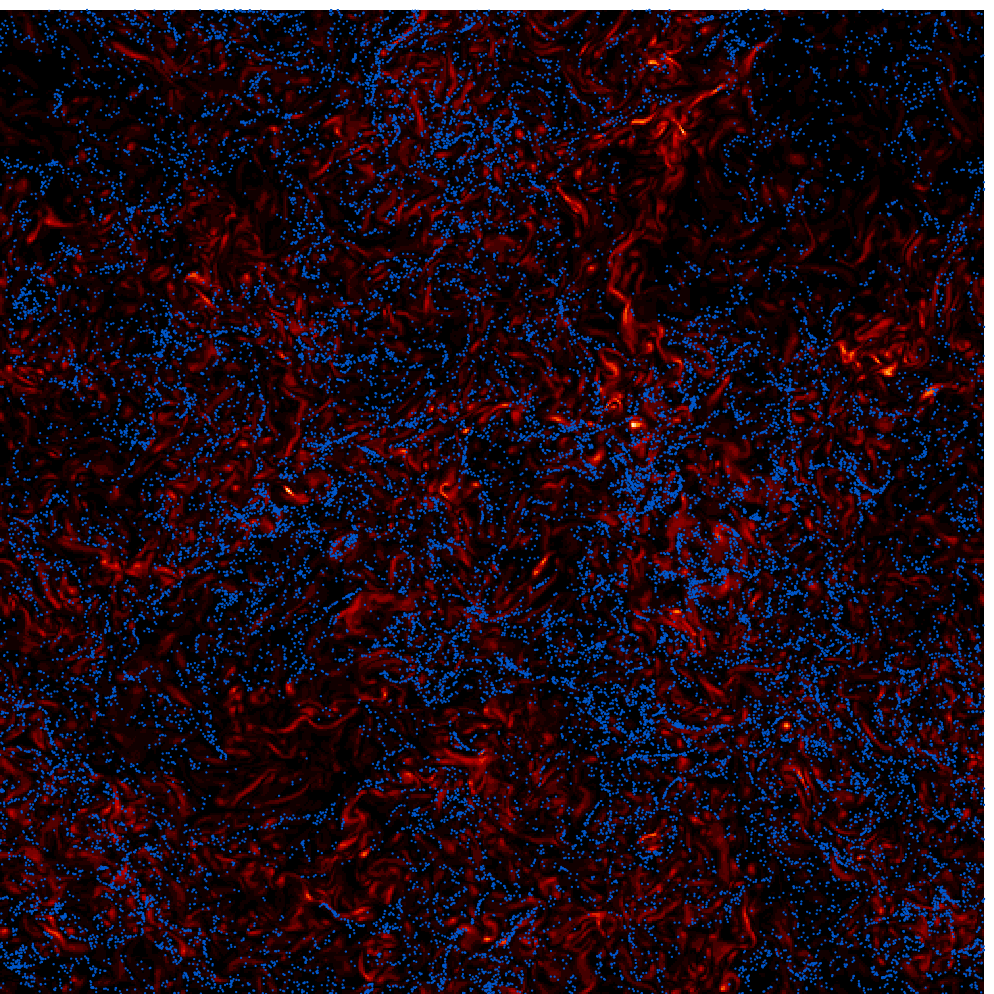}
\includegraphics[width=1.0\columnwidth]{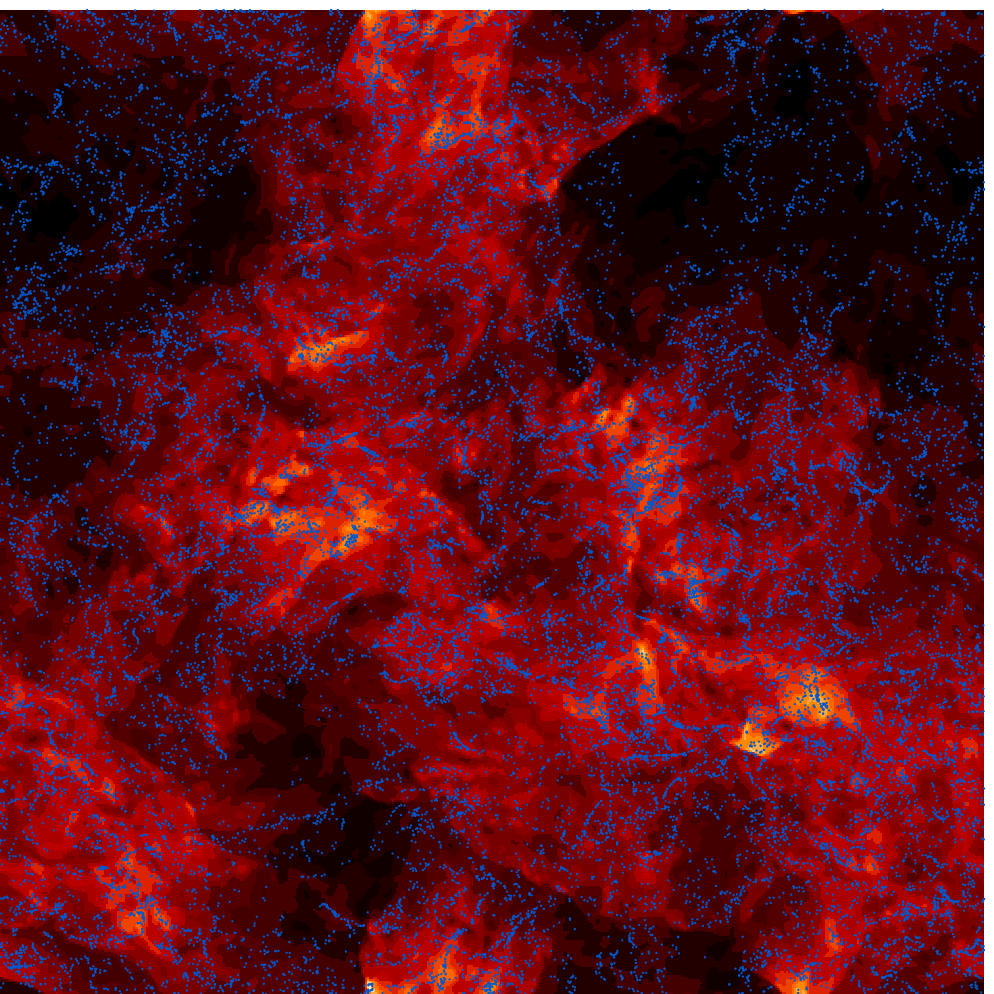}
}
\caption[]{Flow vorticity (left panel) and density(right panel) on 
a slice of the simulation box. The thickness of the slice is  
two computational zones. The color scale is linear 
with vorticity or density, and the red color represents high 
vortcity or density values. Blue dots are locations of particles 
with $St=1.2$.  A clear anti-correaltion is seen between the 
vorticity field and the particle positions, whereas the particle 
distribution is independent of the flow density. The total 
number of particles is the same in the two panels. The 
impression that the left panel has more particles than in 
the right panel is due to the color contrast.}
\label{vortdens}
\end{figure*}  

In our transonic flow, we find that particles are clustered 
with statistical properties very similar to those in incompressible 
flows. The particle clustering found here is not due 
to the compressibility of the gas flow, because very 
strong particle concentration enhancement exists after 
compensating the flow compressibility by dividing the 
particle number density by the flow density. The strongest 
clustering is indeed found for particles with $St\sim 1$. 
Much smaller particles (with much shorter friction timescale) 
behave essentially like tracer particles, and do not show any 
clustering relative to the gas. Clustering of larger 
particles is also weaker and occurs at larger scales. 
Fig.\ \ref{image} shows the position of all the
particles with $St=1.2$ and $St=4.9$ within 
a slice of  thickness equal to 2\% of the computational box. 
At large scales, the spatial distribution of the 
$St=4.9$ particles (right panel) appears to roughly
coincide with that of the $St=1.2$ particles 
(left panel). However, the largest particle densities 
achieved by the $St=1.2$ particles are much larger than those of the 
$St=4.9$ particles (this cannot be fully appreciated in Fig.\ \ref{image}, 
due to the overlap of the particle positions in the densest regions). 
Furthermore, the $St=1.2$ particles show much more small-scale structure
than the $St=4.9$ particles. 

The largest particle densities are found in very elongated structures, 
especially in the case of the $St=1.2$ particles (see Fig.\ \ref{image}). 
The length of these dense particle filaments approaches the size 
of the integral length, $L_1$, of the flow, which is estimated to be 
approximately 0.2 times the simulation box size. The integral scale 
is defined as $L_1 = 3\pi \int k^{-1} E(k) dk/(4 \int E(k) dk) $ 
where $E(k)$ is the energy spectrum. The particle distribution of Fig.\ \ref{image} 
is also characterized by large voids, with sizes spanning the whole inertial 
range up to the $L_1$. The statistics of inertial-range-size voids has 
been studied by Yoshimoto and Goto (2007). The consequences 
of such dense filaments and voids in the particle distribution 
have never been studied in the astrophysical literature. 
We will focus on this important feature of turbulent clustering 
in a separate work.

In Fig.\ \ref{vortdens}, we plot the flow vorticity and density on a thin 
slice of the simulation box, with thickness equal to two 
computational zones, or 5$\eta$. The particle positions are also shown (blue dots). 
From the left panel, we see that particles are mainly 
located in between regions with strong vorticity. This is consistent with 
our physical discussion that inertial particles are expelled by vortices 
and accumulate in the strain-dominated regions.  On the other hand, 
the particle distribution is generally independent of the flow density, suggesting 
that particle clustering in our flows are not caused by or significantly 
affected by compressible modes in our flow.

\section{Clustering Statistics of Identical Particles}

\subsection{The Radial Distribution Function}

\begin{figure*}
\centerline{
\includegraphics[width=1.0\columnwidth]{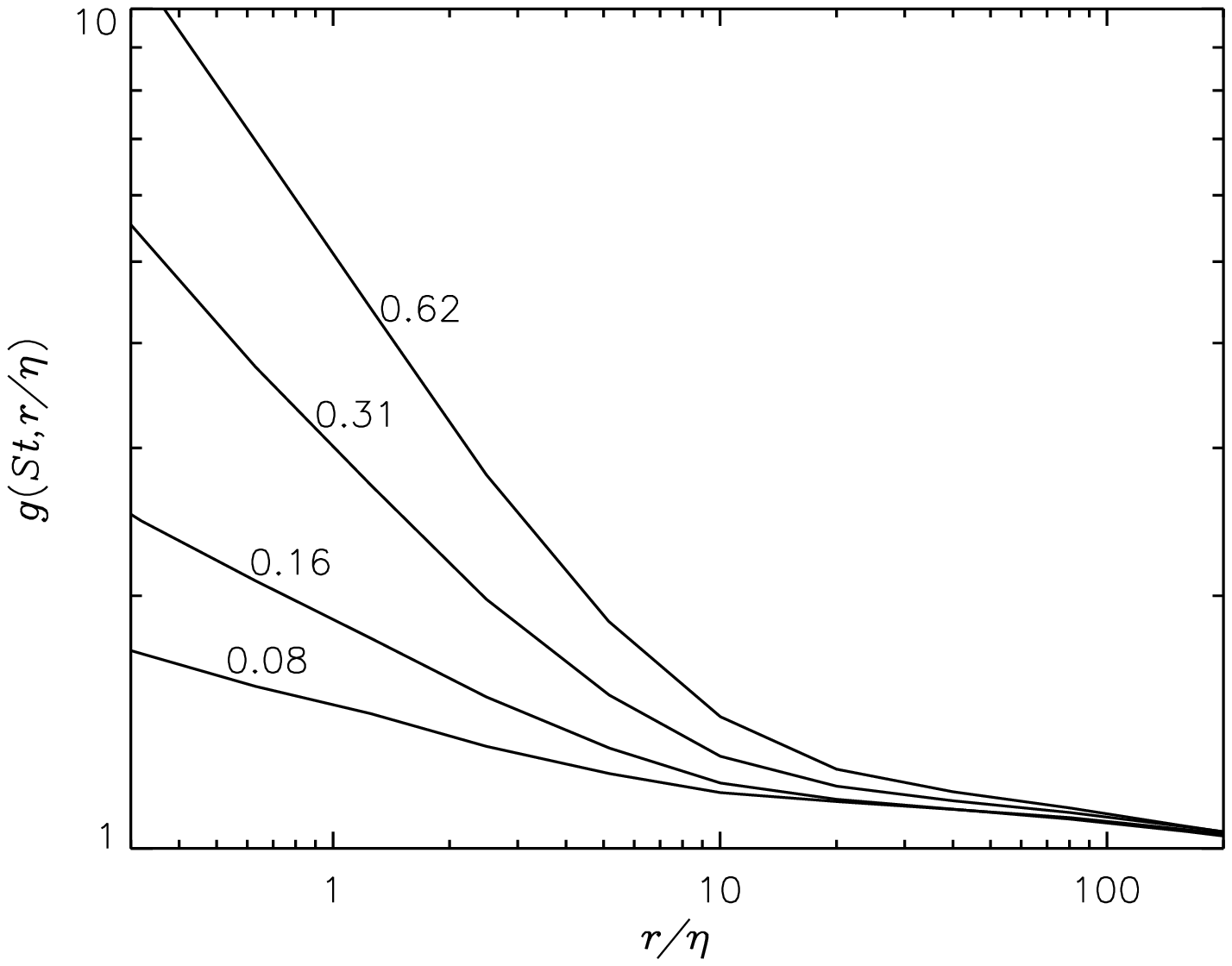}
\includegraphics[width=1.0\columnwidth]{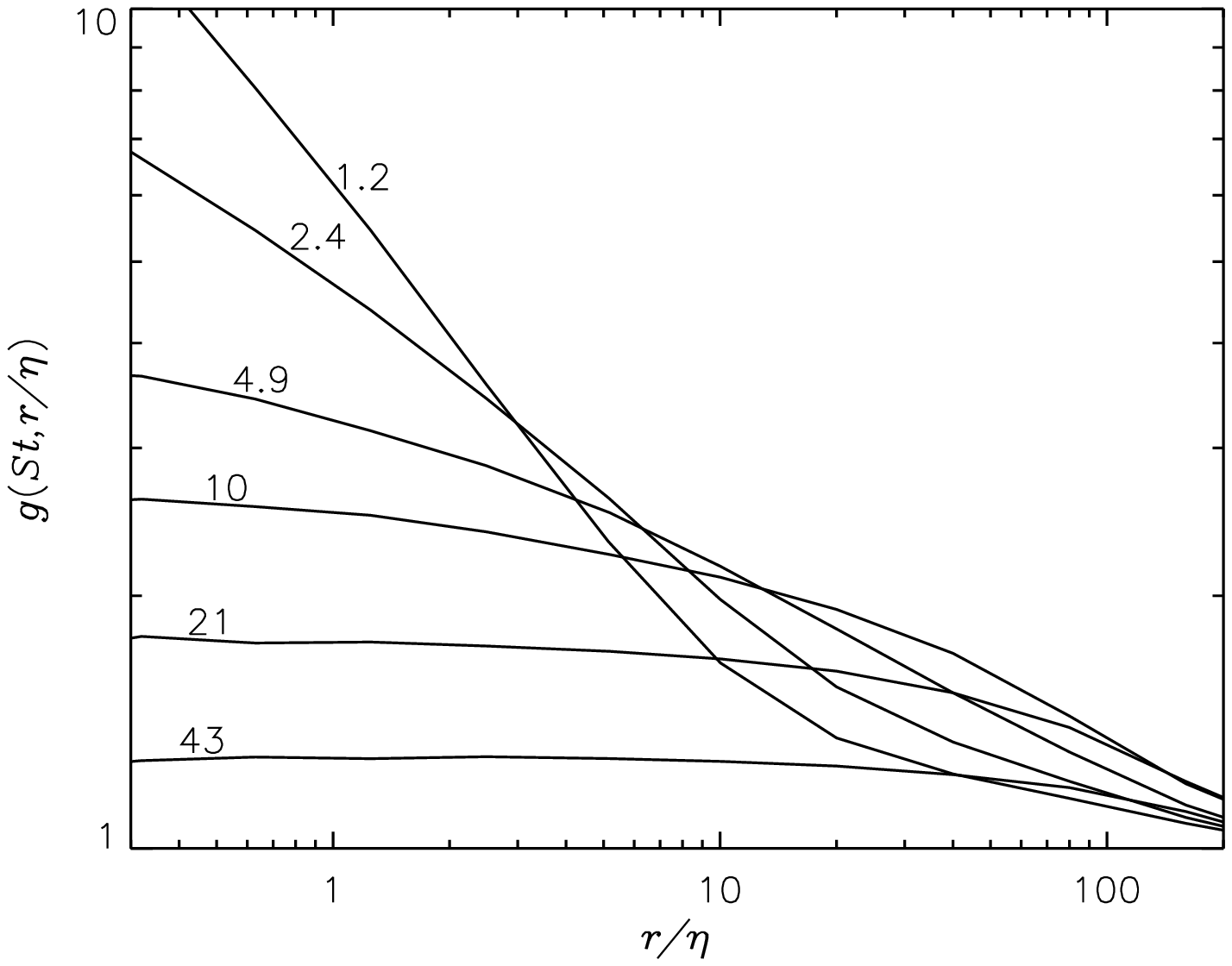}
}
\caption[]{Radial distribution functions for particles 
with $St < 1$ (left panel) and $St > 1$ (right panel). 
The Stokes number for each curve is indicated by a 
nearby label.}
\label{monordf}
\end{figure*}

\begin{figure*}
\centerline{
\includegraphics[width=1.0\columnwidth]{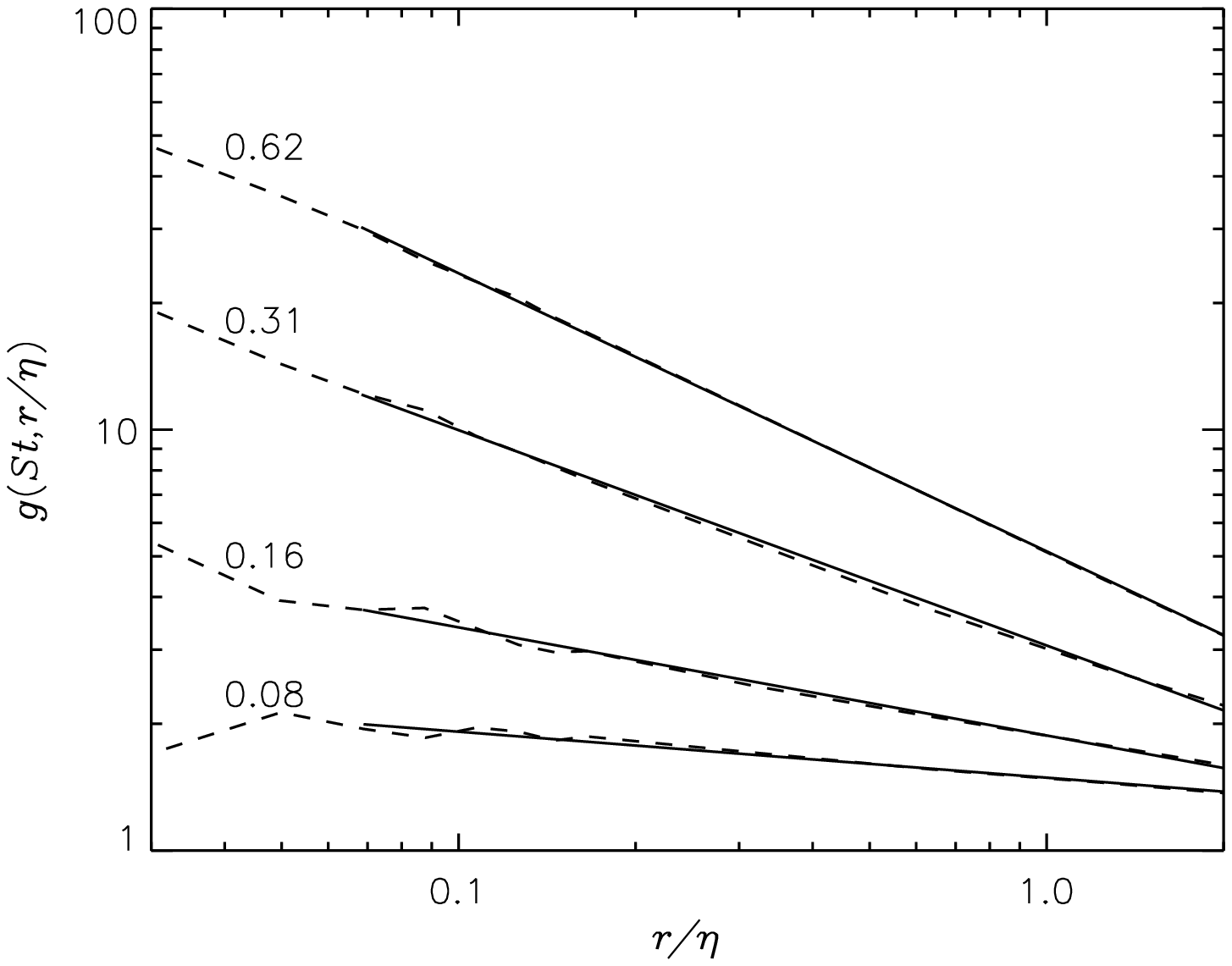}
\includegraphics[width=1.0\columnwidth]{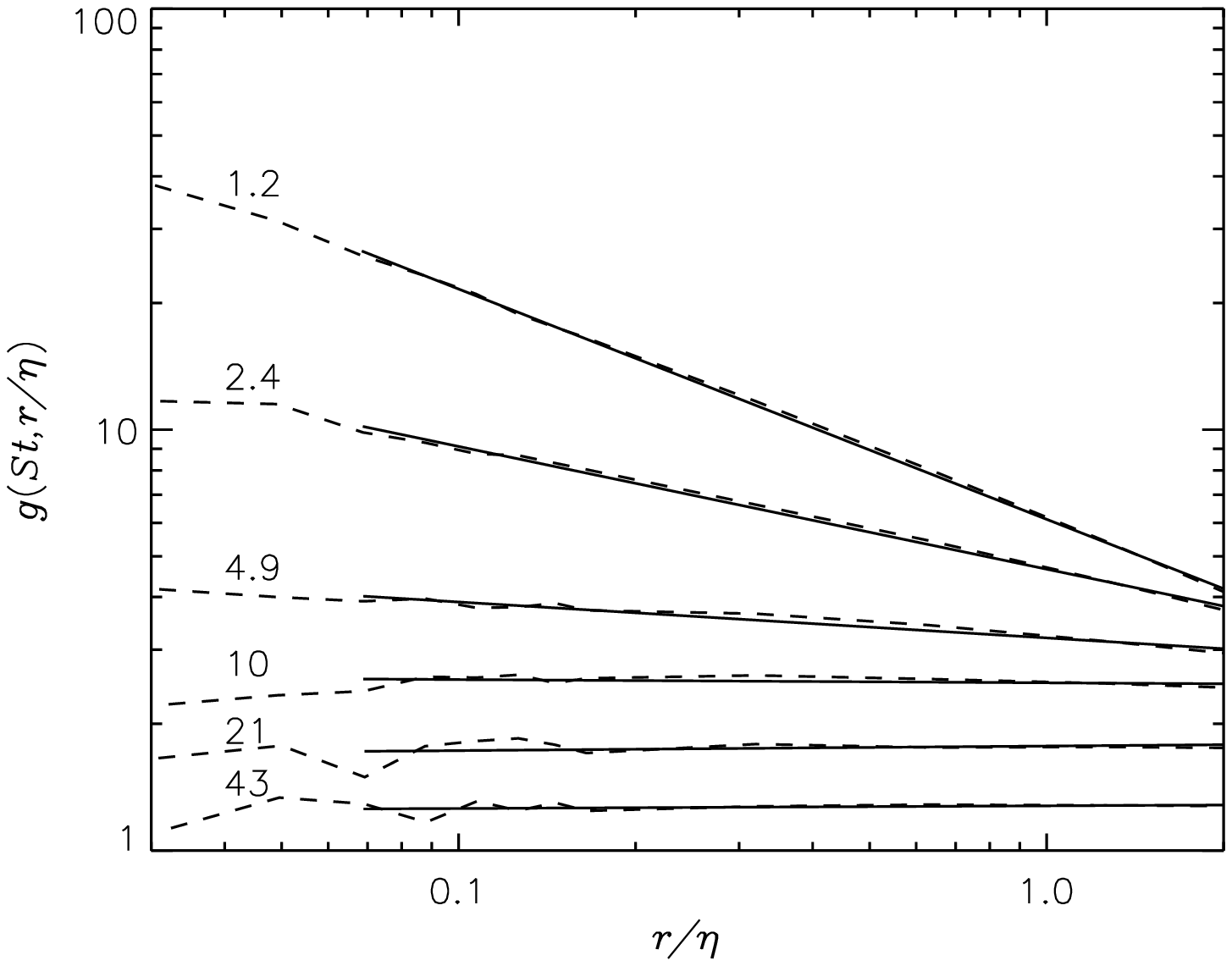}
}
\caption[]{Radial distribution functions at scales 
$r \lsim \eta$. Solid lines are power-law fits. Left and 
right panels correspond to $St < 1$ and $St >1$, 
respectively. The Stokes number for each curve is 
indicated by a nearby label.}
\label{monordfsub}
\end{figure*}

The spatial distribution of particles can be studied by 
computing the density correlation function from the 
particle number density field, $n({\bs x})$. 
The correlation function is defined as, 
\begin{equation}
\xi (St, r) = \frac{1}{\bar{n}^2} \langle (n({\bs x})- 
\bar{n}) (n({\bs x} + {\bs r}) -\bar{n}) \rangle   
\label{corfunc}
\end{equation}
where  $\bar{n}$ is the average particle 
number density, 
and $\langle \cdot \cdot \cdot \rangle$ denotes the 
ensemble average. Alternatively, one can examine 
the fluctuations in the particle density, $n_{\rm r}$, 
coarse-grained over different length scales, $r$ (see, e.g., 
Falkovich and Pumir 2004). The variance, 
$\langle (\delta n_{\rm r})^2 \rangle$ 
(where $\delta n_{\rm r} \equiv n_{\rm r}- \bar{n}$), 
of the coarse-grained density field as a function of $r$ 
provides equivalent statistical information as the 
correlation function $\xi (St, r)$. The two measures 
can be converted from each other using the 
correlation-fluctuation theorem (see below).

Here we use another approach based on the counting 
of particle pairs at given separations. We compute the 
radial distribution function (RDF hereafter), $g(St, r)$. 
It is defined such that the average number, $P(St, r)$, of 
particles in a volume element, $dV$, at a distance, $r(>0)$, 
from a reference particle is given by, 
\begin{equation}
P(St, r)=\bar{n} g(St, r) dV. 
\label{pair}
\end{equation}
This definition is essentially the same as that of 
the two-point correlation function of galaxies in 
cosmology (e.g., Peebles 1980). Clearly, 
for a uniform distribution, $g(St, r)=1$. From their 
definitions, the RDF is equivalent to the density 
correlation function, i.e., $g(St, r) = 1 + \xi(St, r)$ (see 
Shaw 2003). The measured RDF can be used to calculate 
the variance, $\langle (\delta n_{\rm r})^2 \rangle$, 
of the particle density fluctuations at a given 
scale $r$ through the correlation-fluctuation 
theorem. The theorem states that 
\begin{equation}
\frac{\langle (\delta n_{\rm r})^2 \rangle}{\bar{n}^2} = 
\frac{1}{\bar{n} V(r) } + \frac{1}{V (r)} \int_{V(r)} (g(St,r')-1)d^3 r'   
\end{equation}
where $V(r)$ is a volume of size $r$. The derivation 
of eq.\ (7) can be found in, e.g., Landau \& Lifshitz 
(1980) and Peebles (1980). The first term on the r.h.s. is 
the reciprocal of the average particle number in 
$V(r)$ and corresponds to the effect of shot noise (Poisson process). 
The term is negligible for the case of dust particles at length scales, 
$r$, of astrophysical interest. If $\xi(St, r)$ or $g(St, r)$ is a power 
law function of $r$, as found to be the case at $r < \eta$ (see below), 
we have $\langle (\delta n_{\rm r})^2 \rangle \simeq \bar{n}^2 \xi(St, r)$.  

The RDF is especially useful in estimating the collision kernel 
for particle coagulation models. The kernel is  proportional 
to $\bar{n}^2 g(St, d_{\rm p}) \delta v(St, d_{\rm p})$, 
where $\delta v(St, d_{\rm p})$ is the particle relative speed at 
a distance of the particle diameter $d_{\rm p} = 2 a_{\rm p}$ 
(Wang et al.\ 2000).  Both clustering and turbulence-induced 
relative speed tend to increase the particle collision rate.  

For our simulation data, computing the RDF is a better 
approach than the statistical measures based on the 
particle density. This is because the number of particles in 
our simulations is limited, and, at small scales, the 
particle density may not be evaluated with high accuracy. 
On the other hand, we find that the number of particles is 
enough to provide sufficient statistics for the RDF well below 
the Kolmogorov scale, $\eta$ (see Fig.\ \ref{monordfsub}).

\begin{figure}
\centerline{
\includegraphics[width=1.12\columnwidth]{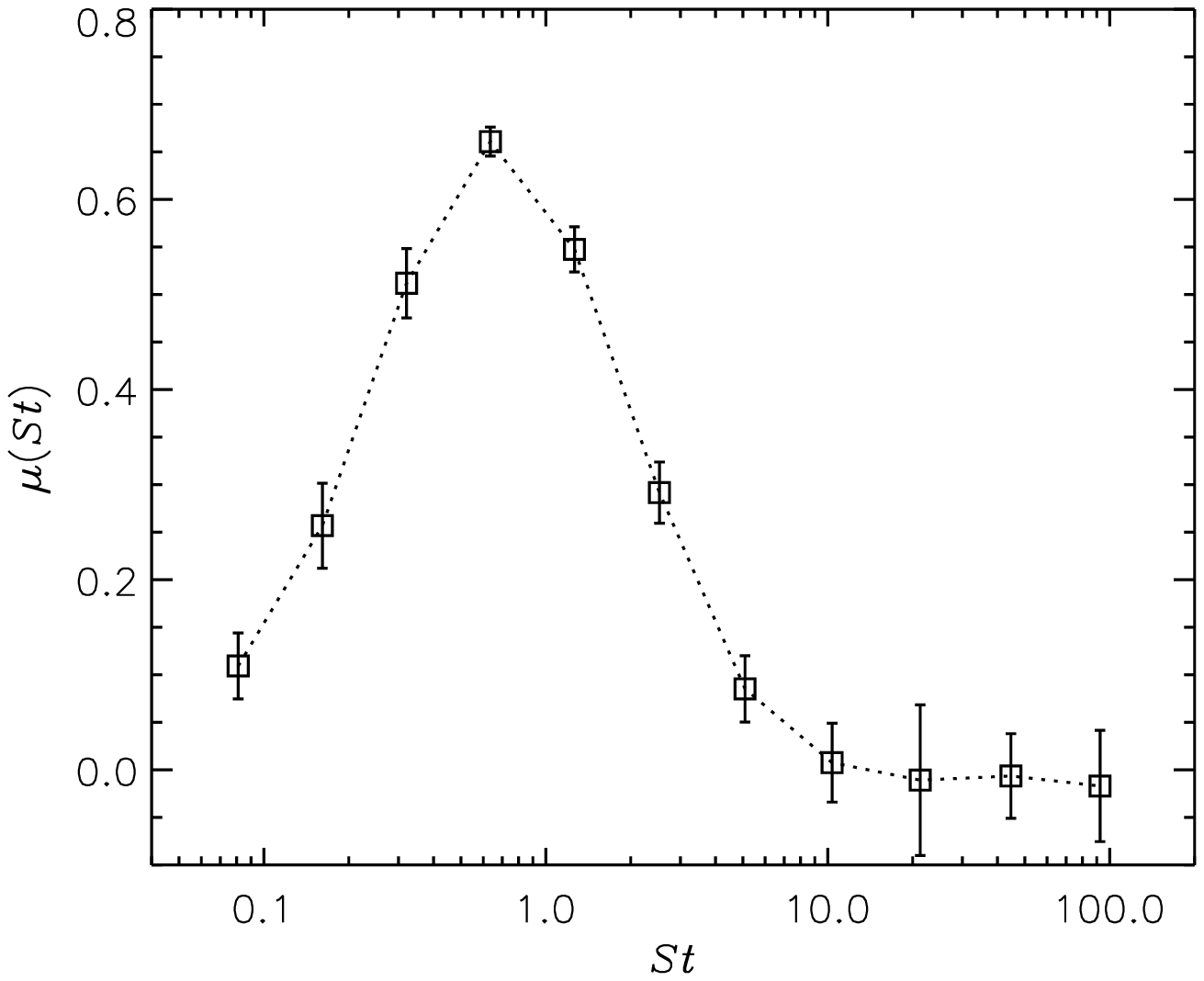}
}
\caption{Scaling exponent, $\mu (St)$, of the RDFs in 
the dissipative range. Error bars show the measurement uncertainty 
($\pm 3 \sigma$).} 
\label{mu}
\end{figure}

Fig.\ \ref{monordf} shows our numerical results for the 
radial distribution function, $g(St, \tilde{r})$,  as 
a function of the particle separation normalized 
to the Kolmogorov scale, $\tilde{r} \equiv{r/\eta}$, for 
different Stokes numbers. The left panel plots 
the RDFs for $St=0.08$, 0.16, 0.31, 0.62 
from bottom to top. The RDF increases 
with the Stokes number at all scales, and this monotonic 
increase actually continues to $St=1.2$ (in the right panel).  
This is in agreement with our 
discussions in \S 2.1 for $St \lsim 1$. For these small Stokes 
numbers, strong clustering is observed at small scales. Consistent 
with previous simulation results, we find that, for 
$r \lsim \eta$, $g(St, \tilde{r})$ can be well fit by a power law,
\begin{equation} 
g(St, \tilde{r}) = C(St) \tilde{r}^{-\mu(St)}. 
\label{eqmu}
\end{equation} 
This is shown in left panel of Fig.\ \ref{monordfsub} where we 
give the power-law fits (solid lines) to the measured RDF 
(dashed lines) in the scale range from 0.03 to 2 $\eta$. 
The exponent $\mu(St)$ increases with $St$ 
for $St \lsim 1$ (see Fig.\ \ref{mu}).  The power-law RDFs 
at $r  \lsim \eta$ suggest self-similarity of the particle 
structures in the dissipation range. On the other hand, 
at scales $ r \gsim 2 \eta$, the RDFs cannot be fit by 
power laws, meaning that the particle density structures are not self-similar 
in the inertial range. The curvature of the RDF curves in the inertial 
range indicates that the clustering process becomes faster and 
faster as the length scale decreases toward the Kolmogorov scale.  
The same trend is seen in the $St=1.2$ curve in the right  
panel.  

\begin{figure}
\centerline{
\includegraphics[width=1.0\columnwidth]{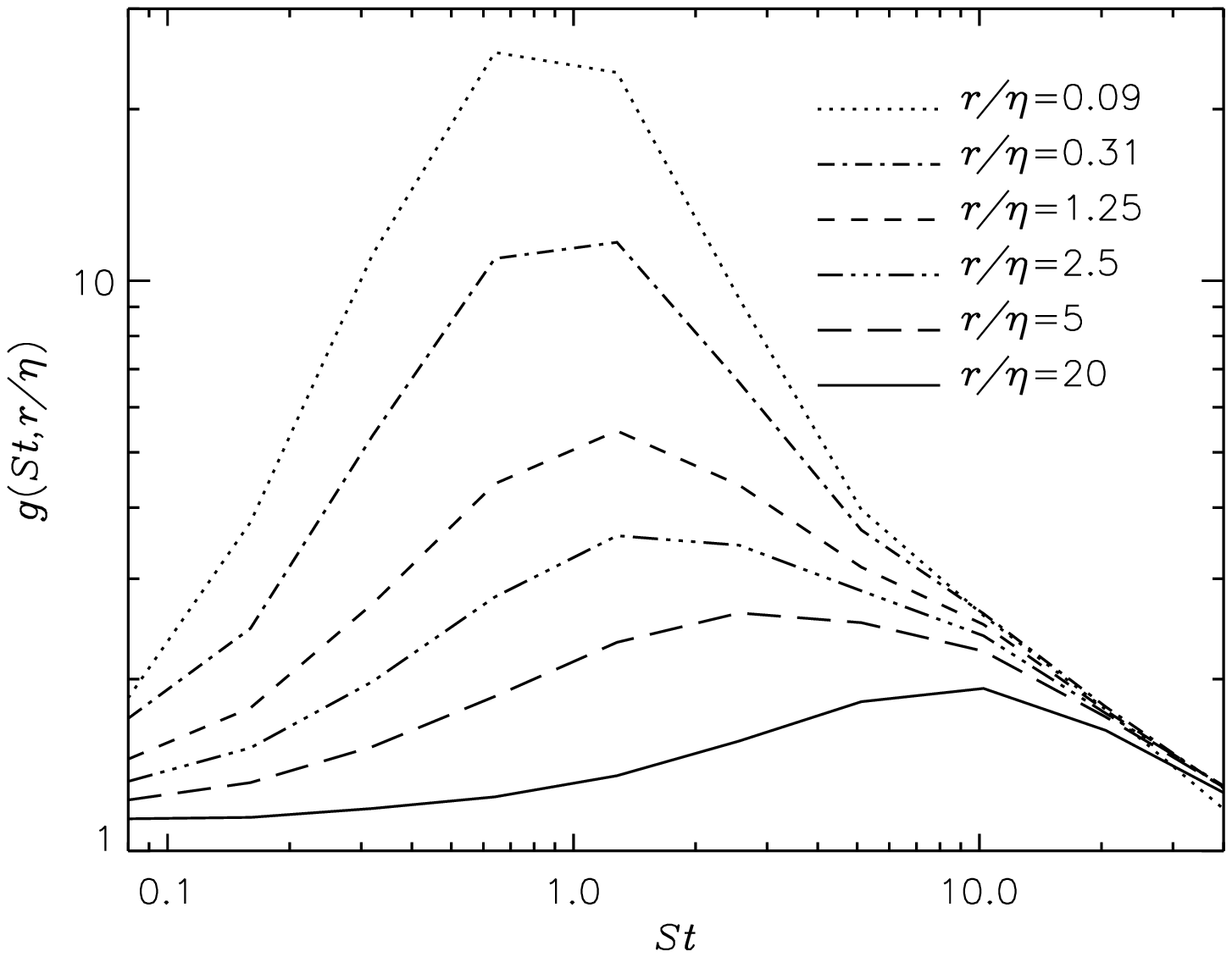}
}
\caption{RDF as a function of the Stokes number at different length scales. }
\label{monordfscales}
\end{figure}

The clustering behavior for $St \gsim 1$ are shown 
in the right panel of Fig.\ \ref{monordf}. From top to bottom, 
the solid lines correspond to $St= 1.2$, 2.4, 4.9, 10, 21 
and 43. The shape of the RDF for $St =1.2$ is very similar to 
those shown in the left panel. However, 
starting from $St=4.9$, the curvature of the RDFs is 
completely different.  For these large particles, the RDF 
first increases steadily toward smaller $r$. As $r$ 
decreases further, a clear decrease in the 
RDF slope occurs at an inertial-range scale for $St \ge 4.9$. 
The scale at which the RDF starts to flatten increases with 
the Stokes number.  Below that scale, the RDF becomes 
essentially flat for $St \gsim 10$, suggesting that no 
significant particle density fluctuations exist at these 
scales. This is in agreement with our physical discussion.  
In \S 2.2, we argued that large particles 
cluster mainly at a scale, $l_{\tau_{\rm p}}$, 
and, below $l_{\tau_{\rm p}}$, the particle relative 
motions are random, and no further clustering occurs. 
This explains the flat part in the RDFs of $St \gg1$ particles. 
Therefore, the scale at which the RDF flattens 
corresponds to $l_{\tau_{\rm p}}$, which increases 
with $\tau_{\rm p}$ as $\tau_{\rm p}^{3/2}$ for $\tau_{\rm p}$ 
in the inertial range. This predicts that the scale for the RDF 
slope change goes like $St^{3/2}$. Unfortunately, due to 
the limited numerical resolution, it is not clear if this 
scaling is strictly obeyed.  In the inertial range, the clustering 
intensity of large particles can be significantly larger than that of small 
particles with $St \simeq 1$, as can be seen from a comparison 
of the $St=1.2$ RDF to the $St= 4.9$, $10$, $21$ 
curves in the scale range from $5 \eta$ to $50 \eta$. The study of 
turbulent clustering at inertial-range scales of astrophysical systems 
should thus pay particular attention to large particles with $St \gg 1$.

In the right panel of Fig.\ \ref{monordfsub}, we plot 
the RDFs in the scale range from 0.03 to 2$\eta$ for $St>1$ 
particles. As in the $St<1$ case, they can also be fit by 
power-laws (solid lines). The RDF slope decreases 
with $St$ in this Stokes number range, and the 
RDF for $St \ge 10$ is completely flat. 
In Fig.\ \ref{mu}, we show the scaling exponent $\mu (St)$ 
as a  function of $St$, which peaks at $St=0.63$.

In Fig.\ \ref{monordfscales}, we plot the RDF as a function of the Stokes 
number at six length scales. The RDF decreases with 
increasing length scale. At scales below $\eta$, the degree 
of clustering strongly peaks at $St \sim 1$.  The 
peak systematically moves to larger Stokes numbers as 
the length scale $r$ becomes larger than $\eta$. As discussed above, the 
strongest clustering at the inertial-range scales is 
from particles with $\tau_{\rm p}$ corresponding 
to the inertial range.  

Our results for the RDF are in good agreement with 
Collins and Kesiwani (2004), who investigated
clustering of $St \sim 1$ particles 
in incompressible flows using DNS. 
Fig.\ 5 of Collins and Kesiwani (2004) 
shows the exponent $\mu$ measured at 
different resolutions. The exponent has apparently 
converged at their highest resolution ($192^3$). 
At that resolution, $\mu$ is around 0.69 for 
$St = 0.4$ and 0.7, and decreases to 0.65 
and 0.50 at $St=1$ and $St=1.5$, respectively. 
These $\mu$ values match very well with 
our Fig.\ \ref{mu}. The agreement provides an important support 
for the numerical schemes adopted in our study, 
including the interpolation method for the velocity field  at sub-grid scales. 
It also suggests that our simulation results can be 
reliably used to explore the clustering statistics in 
essentially incompressible flows. The coefficient 
$C(St)$ in Eq.\ \ref{eqmu} from our simulations is 
also consistent with Collins and Keswani (2004). 
The coefficient is equal to the RDF at $r=\eta$, and from the
$St=1.2$ curve in our Fig.\ \ref{monordf} we have $C \simeq 6.2$ for $St=1.2$. 
This is close  to the measured values  ($\sim 6-7$) of $C(St)$ 
at $St \simeq 1$ from the 192$^3$ run of 
Collins and Keswani (2004). The agreement also has 
interesting implications for the Reynolds number dependence of the RDF (\S 4.3.1).

From Fig.\ \ref{monordf}, we see extremely strong clustering at 
very small scales for particles with $St \sim 1$. The RDF 
keeps increasing with decreasing length scale 
below $\eta$. 
From the RDF plot at smaller scales (Fig.\ \ref{monordfsub}), 
we find that the RDF is as large as $50$ at 
$r \simeq 0.03 \eta$ for $St = 0.62$. This 
indicates very strong clustering: the probability 
of finding another particle across a small distance 
to a given particle can be enhanced by a factor 
of $\sim 100$, relative to the case of uniformly distributed 
particles. The rms concentration, $\langle \delta n_{\rm r}^2\rangle^{1/2}/\bar{n}$, 
at this scale is very large, $\sim10$.

Particle clustering at small scales can strongly enhance the 
particle collision rates. This needs to be accounted for in particle 
coagulation models. As mentioned earlier, the collision rate  
is proportional to the RDF, $g(St, d_{\rm p})$, 
at a separation equal to the particle diameter $d_{\rm p}$. 
The collision frequency is thus $g(St, d_{\rm p})$ times larger than 
if turbulent clustering is neglected. In other words, 
turbulent clustering reduces the coagulation/collision 
timescale by a factor of  $g(St, d_{\rm p})$.
The particle diameter is usually much smaller than 
$\eta$, and $g(St, d_{\rm p})$, at $d_{\rm p}$ can be 
evaluated by extrapolation using our power-law fits at scales 
below $\eta$.

The increase of the RDF toward the particle size, 
$d_{\rm p}$, may be suppressed by the Brownian motions of 
particles. The Brownian motions diffusively spread the 
particles and tend to smear out the particle 
density fluctuations. There is a scale below which the 
Brownian motions dominate over the production 
of particle fluctuations by turbulent clustering. We will 
refer to this scale as the Brownian scale and denote 
it as  $l_{\rm B}$. We give a derivation of $l_{\rm B}$ in 
Appendix B. Below $l_{\rm B}$  no further clustering is 
expected, and the radial distribution function should be flat. 
Therefore, if the Brownian scale is larger than the particle 
diameter, we have $g(St, d_{\rm p}) \simeq g(St, l_{\rm B})$, 
and the extrapolation should stop at $l_{\rm B}$.
On the other hand, if $l_{\rm B} \lsim d_{\rm p}$,
we need to extrapolate the RDF down to $d_{\rm p}$ 
for the estimate of $g(St, d_{\rm p})$.

In summary, we have measured the RDF for particles of 
different sizes from our simulation data, and the results are 
consistent with the physical discussions in \S 2.  Strongest 
clustering are found to occur at $St \sim 1$. The RDFs in 
the dissipation-range scales follow power laws and the 
exponent $\mu(St)$ is largest at $St \simeq 1$. The 
power-law increase of the RDF toward small scales 
implies a strong effect of turbulent clustering on the 
particle collision rate. Large particles ($St>1$) cluster 
primarily at inertial-range scales, where their clustering 
intensity is larger than that of $St<1$ particles.  

\subsection{The Particle Concentration PDF}

As a second order statistical measure, the RDF 
reflects the rms amplitude of the particle density 
fluctuations. In some applications, high-order 
statistics, corresponding to clusters with extreme 
particle density, are of particular interest. For 
example, in \S 6 we will discuss planetesimal 
formation models based on particle 
clusters of high concentration level in 
protoplanetary disks. The probability of 
finding  these dense clusters can be estimated from the 
probability density function (PDF) of the particle 
concentration. 

We will compute the concentration PDF at different 
length scales. At each length scale, $r$, we consider 
regions of size $r$,  and in each region we define a 
particle concentration $C \equiv n_{\rm r}/\bar{n}$ 
where $n_{\rm r}$ is the average number density in that 
region. We denote as $P_{\rm r}(C)$ the concentration 
PDF at  scale $r$, which represents the probability 
of finding clusters of size $r$ with a given particle 
concentration, $C$.  
 
\begin{figure}
\centerline{\includegraphics[width=1.0\columnwidth]{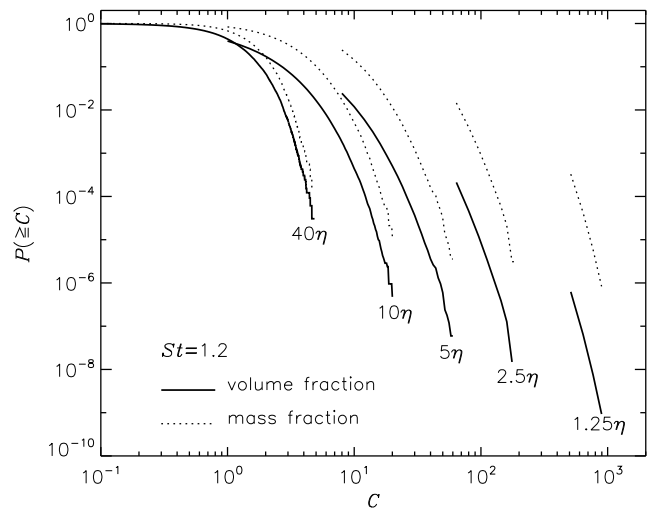}}
\caption{Cumulative PDF of the particle concentration 
for $St=1.2$ particles at different length scales.}
\vspace{0.1in}
\label{monopdf}
\end{figure}

The computation of the PDF from our simulation 
data is done as follows. We first divide the simulation 
box into cubes of size $r$ and evaluate the 
particle number density and the concentration 
in each cube. The particle density (and hence the 
concentration) can be accurately measured only if the 
number of particles in a cube is much larger than 1. We 
thus decided to only count the cubes containing 
4 or more particles, while the cubes with less 
particles were simply ignored. Therefore, 
the measured PDF starts from a minimum 
concentration corresponding to 4 particles per 
cube. The minimum increases with decreasing length scale 
$r$ (see Fig.\ \ref{monopdf}) because, for smaller cube 
sizes, 4 particles per cube implies a larger concentration. 
Using this method, we computed the concentration PDF down to 
the scale $\simeq \eta$. 

Due to the limited number of particles, 
the measured PDFs can be contaminated 
or even dominated by the Poisson noise, 
especially at small scales. We compared 
the measured concentration PDF at each scale 
to the PDF that arises purely from Poisson poise. 
At small scales ($r \lsim 5\eta$), we only 
measured the high tails of the PDF, and the 
probability in the tails appears to be well above the Poisson 
noise PDF (by at least two orders of magnitude) for Stokes 
numbers in the range $0.3 \le St \le 10$. 
Particles outside this range are less clustered, and the
measured PDFs are close to the Poisson PDF. 
For those particles, we need a larger number 
of particles in the simulations to obtain accurate statistics. 
At large scales ($r \gsim 10\eta$), we have good 
measurements  for particles with $0.16 \lsim St \lsim 40$, 
whose PDFs are significantly broader than the 
Poisson PDF. 
   
In Fig.\ \ref{monopdf}, we plot the cumulative PDF,  
$P_{\rm r}(>C) =\int_C^{\infty} P_{\rm r}(C')dC'$ at 
different scales for $St=1.2$. The PDF 
is broader at smaller scales, corresponding to 
the increase of the RDF with decreasing 
length scale at $St \sim 1$.  As $r$  decreases 
toward $\eta$, the broadening of the PDF appears 
to be faster, consistent with the trend observed in 
the RDF for $St=1.2$. The scale dependence 
here is quite sensitive. The PDFs at large 
scales ($r \gg \eta$) are much narrower than those at 
small scales. 

From Fig.\ \ref{monopdf}, we see that at $1.25 \eta$ 
there is a finite probability of finding regions 
with very high concentration enhancement, $C \sim10^3$. 
The trend that the PDF becomes broader with 
decreasing scale suggests that even higher 
density clusters may be found at scales below 
$\sim \eta$. The growth of the PDF tail may continue to the Brownian 
scale, below which further clustering is suppressed. 
However, the PDFs shown in Fig.\ \ref{monopdf} do not 
account for the back reaction from the particles to 
the carrier flow, which is not included in our simulations. 
The back reaction cannot be neglected in regions with 
$C \gsim 10^3 $, because the local mass loading factor 
$\Phi_{\rm m}$ is much larger than 1 (assuming an 
average dust-to-gas ratio of 0.01). Therefore, the 
back-reaction may significantly affect the high tails 
of $P(C)$ (Hogan and Cuzzi 2007). This will be discussed 
in more details in \S 4.5.

Following Hogan et al.\ (1999), we also considered the 
PDF with mass-weighting, $P_{\rm m} (C)$, which is related 
to the volume-weighted PDF, $P(C)$, by $P_{\rm m} (C) = C P(C) /\langle C \rangle $ 
(here the subscript ``r" for the scale dependence is 
dropped for simplicity of the notation).  
The cumulative PDF with mass-weighting is thus 
$P_{\rm m}(>C)=\int_C^{\infty}C' P(C')dC' /\langle C \rangle$, 
which is the fraction of the total number 
(or mass) of particles experiencing a concentration 
larger than $C$. The cumulative mass-weighted 
PDF is plot as dotted lines in Fig.\ \ref{monopdf}.  
We find that that the volume- and mass- weighted PDF tails at $r =2.5 \eta$  
in our Fig.\ \ref{monopdf} are quite close to the results  
in Hogan et al.\ (1999) (their Fig.\ 3c and 3d, 
respectively) for $St=1$ particles at $r=2\eta$ in a  
$Re_{\lambda} = 140$ flow.  The cumulative PDF 
with mass-weighting has much broader tails. For 
example, the PDF tails at $r=\eta$ in our 
Fig.\ \ref{monopdf} show that the mass-weighted 
probability for $C \gsim 10^3$ is about $10^3$ times 
larger than the volume-weighted one. We note that the 
PDFs shown in Fig.\ 4 of Cuzzi et al.\ (2001) 
and in Fig.\ 1 of Cuzzi et al.\ (2008) correspond 
to the mass-weighted cumulative PDFs in Fig.\ 3d 
of Hogan et al.\ (1999).

Although our Fig.\ \ref{monopdf} looks similar to  Fig.\ 1 in Cuzzi 
et al.\ (2008), they are different. In our figure, the particle size and 
numerical resolution are fixed, the curves correspond to 
different length scales. On the other hand, Fig.\ 1 of Cuzzi et
 al.\ (2008) shows the concentration PDFs at different 
numerical resolutions with the Stokes number and the 
normalized length scale fixed at $St \simeq 1$ and $\tilde{r} =2$ 
respectively.

\begin{figure}
\centerline{\includegraphics[width=1.0\columnwidth]{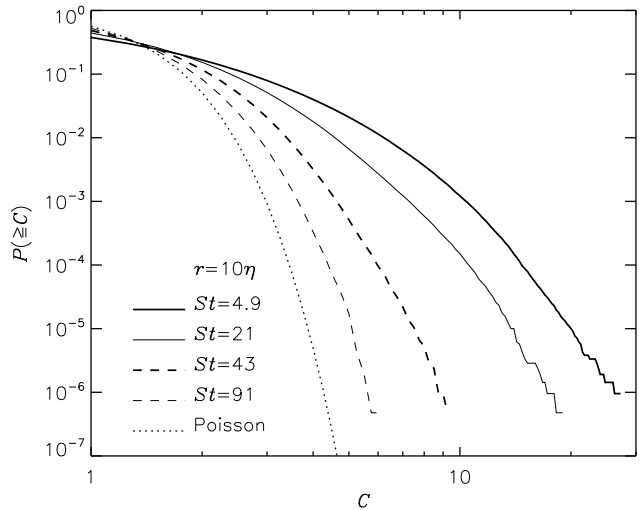}}
\caption{Cumulative PDF of the particle concentration at $r=10\eta$. 
At this scale the PDF width peaks at $St=4.9$. For larger $St$, 
the PDF becomes narrower. The dotted line shows the PDF of 
Poisson noise.}
\label{monpdfstokes}
\end{figure}

We also computed the concentration PDFs for other 
Stokes numbers. At a given scale, the PDF tails 
as a function of $St$ have a similar behavior 
as the RDFs shown in Fig.\ \ref{monordfscales}.  
At $r \simeq \eta$, the PDF tail first broadens with 
increasing $St$, and reaches a maximum width 
at $St \simeq 1$.  As $St$ increases further, the PDF 
tail becomes narrower.  Also consistent with the RDF in Fig.\ \ref{monordfscales},  
the Stokes number at which the PDF width reaches 
maximum becomes larger with increasing 
length scale. For example,  at $r=10 \eta$ and $40 \eta$, the PDF 
tail reaches maximum at $St= 4.9$ and $10$, respectively. 
Again the highest clustering intensity at inertial-range scales is from particles 
with $\tau_{\rm p}$ in the inertial range. In Fig.\ \ref{monpdfstokes}, we show 
the dependence of the PDF tail on $St$ at the scale $r=10\eta$.  
Starting from $St=4.9$ where the PDF has the maximum width, 
the tail becomes narrower as $St$ increases. At $St \gsim 93$, 
the PDF is quite close to the Poisson PDF, indicating only 
a slight or negligible clustering effect.

\subsection{Reynolds number dependence}

Currently available numerical studies are 
far from resolving scales around the turbulence 
dissipation scale, $\eta$, in interstellar clouds 
or protoplanetary disks, as the characteristic 
Reynolds number in these astrophysical 
systems is $Re \gsim 10^6$. The possible 
dependence of the clustering properties 
on the Reynolds number must be carefully 
examined, if results of numerical simulations 
are to be applied to astrophysical environments. 

The $Re$ dependence is usually discussed 
in a unit system where the length scale and 
the particle friction timescale are normalized, 
respectively, to the Kolmogorov length and 
time scales in the carrier flow. Numerical 
simulations used to study the $Re$-dependence 
usually keep the large-scale properties (such as the rms 
velocity and the integral length scale) roughly constant, 
and decrease the viscosity (and hence the Kolmogorov scale) 
with increasing resolution. In the statistical analysis, 
these studies normalize all the quantities to the 
smallest scales (i.e., $\eta$ and $\tau_{\eta}$) in the 
simulated flows, and examine how the clustering properties at 
given $St$ and $\tilde{r}$($\equiv r/\eta$) change with the 
Reynolds number. Note that, in the comparison of 
the clustering statistics in simulated flows with 
the same large-scale properties, but different $Re$, 
given values of $St$ and $\tilde{r}$ correspond to 
different particle sizes (i.e., different $\tau_{\rm p}$) 
and different actual length scales, $r$ (larger 
particle size and length scale in the lower $Re$ 
flow).

The Reynolds number dependence of particle clustering 
has been discussed in a number of numerical studies using 
simulations at different resolutions (e.g., Hogan, Cuzzi and 
Dobrovolskis 1999, Wang et al.\ 2000, Reade and Collins 
2000a, Hogan and Cuzzi 2001, Falkovich and Pumir 2004, 
Collins and Keswani 2004). Here we give a brief summary 
of the results from these studies. 
 
\subsubsection{Reynolds number Dependence of the RDF} 

The RDF was found to increase with $Re$ at very low $Re$.  
Wang et al.\ (2000) computed the RDF at $r=\eta$ in four 
simulated flows with the Taylor Reynolds number, $Re_{\lambda}$, 
in the range from 24 and 75. Their results show that the RDF 
increases linearly with $Re_{\lambda}$ (i.e., by a factor of 3 as 
$Re_{\lambda}$ goes from 24 to 75), and that the shape of 
$g(St, \tilde{r}=1)$ as a function of $St$ does not change with 
$Re_{\lambda}$ (see also Hogan and Cuzzi (2001)). A similar 
increase in the RDF with increasing $Re_\lambda$ is observed 
by Reade and Collins (2000a) for $r=0.025 \eta$ in a similar 
range of $Re_\lambda$.

At higher resolutions, different conclusions were obtained in 
different studies. Using numerical simulations with 
$Re_\lambda \lsim 130$, Falkovich and Pumir (2004) 
found that the scaling exponent, $\mu (St)$, for $St <1$ 
at $r <\eta$ has a significant increase with increasing 
$Re_\lambda$. This dependence has been suggested 
to have important implications for the growth of droplets 
in terrestrial clouds. On the other hand, Collins and Keswani (2004), 
who explored $St \sim 1$ particles in a similar $Re_\lambda$ 
range (up to 152), showed that the scaling exponent, $\mu(St) $, is 
essentially independent of $Re_\lambda$, and that the 
coefficient, $C(St)$, first increases with $Re_\lambda$ 
at small $Re_\lambda$, and then converges to a constant 
at $Re_\lambda =152$. These suggest that the RDF 
may be $Re$ independent at sufficiently large $Re$. 
In \S4.1, we found that  the RDF of $St \simeq 1$ 
particles in our simulated flow with $Re_\lambda \simeq 300$ 
are in good agreement with Collins and Keswani (2004). This 
supports the claim by Collins and Keswani (2004) that the 
RDF is $Re$-independent at high Reynolds numbers. 
However, we think that a conclusive answer to the $Re$ 
dependence of the RDF still needs confirmation 
from simulations of higher-resolutions.  

The $Re$ dependence of the RDF of $St>1$ 
particles in the inertial range has not been 
investigated. These large particles cluster at 
a scale, $l_{\tau_{\rm p}}$, in the inertial range, 
which were barely resolved in existing studies. 
To accurately capture the clustering statistics at 
the scale $l_{\tau_{\rm p}}$, an extended 
separation between $l_{\tau_{\rm p}}$ and 
the low outer scale, $L$, is needed, where the RDF 
increases toward smaller scales (see Fig.\ \ref{monordf}). 
This requires even higher numerical resolutions 
than for the study of $St \lsim 1$ particles. 
We speculate that the $Re$ dependence for $St >1$ particles 
would be weaker than that for $St \lsim 1$ particles. 
In \S 2.1, we showed that, for $St <1$, the divergence 
of the particle flow is proportional to the 
velocity gradient squared, which has a fairly strong 
$Re$ dependence. In contrast, the effective 
compressibility estimated for particles in \S2.2 
does not depend on the flow properties. Therefore 
the $Re$ dependence for $St >1$ particles 
is expected to be weaker. If the RDF of 
$St \lsim 1$ particles is $Re$-independent 
at large $Re$, the same is probably also 
true for $St>1$. Future numerical 
studies can test this speculation.

\subsubsection{Reynolds Number Dependence of the PDF}  

In \S 2.1, we derived the divergence of the 
particle flow for $St \lsim1$, and found that 
the divergence has a quadratic dependence on the flow 
velocity gradient.  The PDF of the velocity gradient 
in a turbulent flow is known to broaden with 
increasing $Re$. The same is thus expected 
for the PDF of the particle flow divergence, 
meaning that the probability of strong compressing 
or expanding events is higher at larger $Re$.  As a 
consequence, the concentration PDF for $St \lsim 1$ 
particles is likely to become broader with increasing 
Reynolds number.  
Note that a $Re$-dependent PDF does not 
suggest that the RDF, a second-order statistical 
measure, must also depend on $Re$. It is 
possible that the tails of a PDF broaden 
considerably with $Re$, while the second order moment is 
constant.

Broadening of the particle concentration 
PDF with increasing resolution was found in 
the numerical study of Hogan et al.\ (1999) 
for particles with $St =1$. Hogan et al.\ 
(1999) carried out a multifractal analysis 
of the particle concentration field that 
can be used to extrapolate the PDFs 
measured from low-$Re$ simulations to 
realistic values of $Re$. They computed the 
singularity spectra of the particle concentration 
field at different scales ($2 \le \tilde{r} \le 8$), in 
simulations with three different values of $Re$. Fig.\ 2 of 
Hogan et al.\ (1999) shows that, at each Reynolds 
number, the spectra are different at different 
scales, indicating that the particle density 
structures are {\it not} self-similar at 
scales above $2\eta$. This is consistent with 
our observation that the RDF is not a power-law at scales 
above $2 \eta$ in our simulations\footnotemark\footnotetext{The singularity 
spectrum at scales $ \lsim \eta$ may
be scale-independent because particle 
structures at these scales appear to 
be self-similar, based on the power-law 
RDFs below $\sim \eta$.}.  
Strictly speaking, the scale dependence 
of the singularity spectrum means that the 
particle structures are not ``fractals".  
However, the multifractal analysis provides 
useful information on how the clustering process 
proceeds with decreasing length scales.     
The singularity indices are significantly 
smaller at smaller scales, suggesting the 
development of strong particle density structures 
becomes faster and faster toward smaller $r$.   

On the other hand, the singularity spectra at a 
given scale, $\tilde{r}$, are found to be independent 
of the Reynolds number. Based on this dependence,  
Hogan et al.\ (1999) gave a model to extrapolate 
the concentration PDF from simulation results 
to that at realistic Reynolds numbers. Applying 
the extrapolation to $Re$ values typical of 
turbulence in planetary disks, Cuzzi  et al.\ (2001) 
found a significant probability of finding 
regions (of size $ \sim \eta$ ) with extreme 
concentration enhancement ($ C \sim 10^4-10^5$).

The singularity spectrum of the particle concentration 
at $2 \eta$ measured by Hogan et al.\ (1999) is very
similar to that of the dissipation rate in the turbulent 
flow (see their Fig.\ 2). The reason is probably that 
the particle velocity divergence has a quadratic 
dependence on the flow velocity gradient, which 
is similar to that of the dissipation rate.

Hogan et al.\ (1999) only investigated particles 
with $St \sim 1$, and the $Re$ dependence of 
the concentration PDF at $St <1$ or $St>1$
has not been studied. We expect that the 
concentration PDF of small particles ($St<1$) 
would broaden with increasing $Re$ in a similar 
way as $St \sim 1$ particles, because the divergence 
of these particles has a similar dependence on the 
velocity gradients.  It is unknown how the 
concentration PDF of $St>1$ particles changes 
with $Re$. As in the RDF case, we argue that, in the 
$St >1$ case, the $Re$ dependence of the 
concentration PDF would be weak in 
comparison to the $St \sim 1$ particles. 
This is again based on our observation in \S 2.2
that the effective compressibility ($\sim 1/\tau_{\rm p}$) 
of large particles at the clustering scale $l_{\tau_{\rm p}}$ 
does not show an explicit dependence on 
the flow velocity gradients or velocity 
differences. 

\subsection{Interpretation of Simulation Results}

Due to the limited numerical resolution, 
the Kolmogorov timescale in simulated 
flows is usually much larger than that in 
a real flow. The Stokes number of a 
particle of a given size would be much 
smaller in a simulation than in the real flow. 
A consequence of this mismatch of the Stokes 
numbers is that the clustering intensity from a 
simulation may not correctly reflect that 
in the real situation, as the clustering statistics 
have a quite sensitive dependence  on $St$. 
Therefore, simulation results involving the 
clustering properties of inertial particles 
need to be interpreted with caution.  

We discuss how the clustering statistics obtained 
in simulations may differ from that in the real flow, 
based on the RDF shown in our Fig.\ \ref{monordfscales}.  
This can be examined from the three correction 
steps given below, which allow us to see how the real RDF 
compares to that from a simulation. We use the subscript ``n" 
to denote the numerical results, and the subscript ``r" for 
the real flow.

First, for a given length scale $r$, $\tilde{r}_{\rm r} $ 
in the real flow is larger than $\tilde{r}_{\rm n}$ in 
the simulated flow. This shifts the RDF in Fig.\ \ref{monordfscales} 
toward lower values of $g$ (larger $\tilde{r}$). Second, the Stokes 
number is larger in the real flow than in the 
simulation. This corresponds to a shift to 
the right side along the RDF curve. If $St_{\rm n}  \ll 1$ 
and the shift moves $St$ closer to unity, this correction could 
give an increase in the clustering strength. 
On the other hand, if $St_{\rm n} \gsim 1$, the shift 
would result in smaller values of $g$. Finally, we 
need to account for the possible Reynolds number 
dependence. The Reynolds number is larger in the 
real flow and, if it exists, the $Re$ dependence 
would move the RDF curves upward (see \S 4.3.1). 
  
We consider a specific example where the 
actual particle size has $St_{\rm r} \gg 1$ 
in the real flow, but by coincidence 
corresponds to $St_{\rm n} \sim 1$ in the 
simulated flow. This example is interesting 
for our discussion on the planetesimal 
formation model in \S 6.3. In this case, 
the first two steps discussed above would 
give a clustering intensity much lower 
than in the simulated flow. In particular, 
the effect of the $St$ correction, is quite strong, 
as the RDF curves in Fig.\ \ref{monordfscales} 
decrease very rapidly with increasing $St$ (for $St>1$). 
Therefore, the RDF measured in the simulation 
would overestimate that in the real flow by 
a large amount, unless there is a  
strong $Re$ dependence. The same 
argument can be made for the width of the concentration 
PDF tails. The $Re$ dependence to 
be applied here is that for the clustering of $St >1$ 
particles at inertial-range scales, which 
has not been studied. In \S 4.3.1 and 4.3.2, 
we argued that the $Re$ dependence for 
these particles is likely to be weak. 
Therefore the $Re$ dependence may not be able to 
compensate the decrease in $g$ resulting 
from the first two corrections.
We thus conclude that, if in a simulation the 
particle Stokes number has an artificial 
value close to unity, the clustering 
intensity of those particles may be 
significantly overestimated. 
This needs to be considered when interpreting results 
from astrophysical simulations.

\subsection{Back Reaction}

We have only considered the effect of the 
turbulent flow on the inertial particles, but 
neglected the dynamical effect of the 
inertial particles on the carrier flow. As shown 
in our simulations, turbulent clustering can 
give rise to regions with particle concentration 
enhanced by a factor of $10^3$ 
(see Fig.\ \ref{monordf}), leading to local 
particle densities even larger than the flow density. 
In these regions, the feedback effect from the 
mass loading is not negligible, 
and a discussion of the two-way interactions 
between the particles and the flow 
is needed. 

The modulation of the carrier flow by the back 
reaction from the particle phase has been shown 
to depend on the particle size. Different results 
have been found for $St<1$ particles and 
$St>1$ ones, concerning how the back 
reaction changes the turbulent kinetic 
energy, how the kinetic energy 
transfers between the flow and particle phases, 
and how the energy spectrum of the flow 
is affected by the two-way coupling 
(Sundaram and Collins 1999, Boivin, Simonin and Squires 1998, 
Ferrante and Elghobashi 2003,  Shotorban and 
Balachandar 2009). 
 
Here we are more interested in the effect of 
the two-way coupling on the clustering intensity. 
From brief discussions in Sundaram and Collins 
(1999) (for $St>1$) and in Shotorban and 
Balachanar (2009) (for $St<1$), we see that including the back 
reaction gives only a slight change ($\lsim10\%$) 
in the RDF and the particle concentration variance. 
It seems that the 2nd order clustering statistics is 
not significantly affected by the back reaction if the rms 
mass loading factor is smaller than 1. 
A systematic study of the effect of two-way coupling 
on the RDF is needed to confirm if this is indeed the case. 
 
On the other hand,  the particle feedback can considerably 
affect the tails of the particle concentration PDF because 
clusters of high concentration levels induce much larger 
mass loading than the average. Hogan and Cuzzi (2007) 
studied the back-reaction effect on the concentration 
PDF for particles with $St =1$. They built up a model 
assuming that the development of the fluctuations 
in the particle concentration and the flow enstrophy 
(defined as vorticity squared, $\omega^2$)
can be described as a joint cascade process. In the model, a 
flow parcel breaks up into two equal-sized subdivisions in each
cascade step, and the partitioning of the particle concentration 
and the flow enstrophy in the two subdivisions is 
controlled by a probability distribution, called the 
multiplier PDF. Hogan and Cuzzi (2007) computed 
the multiplier PDF for the step from $3 \eta$ to $1.5\eta$ 
in their simulations, and found that the multiplier 
PDF becomes narrower as the mass loading factor, 
$\Phi_{\rm m}$, exceeds $\sim 10$. When 
$\Phi_{\rm m}$ becomes larger than $\sim 100$, the 
multiplier PDF is essentially a delta function, meaning 
that the bifurcation of the particle density stops in these 
highly loaded regions. This sets an upper limit 
for the concentration enhancement: the particle 
density cannot exceed 100 times the flow density. 
In short, the back reaction from particles has been 
found to suppress the probability of forming particle 
clusters with extreme concentration enhancements.     
\\

The two-way interactions between the dust particles 
and the turbulent flow give rise to an interesting 
phenomenon in differentially-rotating circumstellar 
disks. Youdin and Goodman (2005) found that, 
with two-way coupling, the presence of a 
radial pressure gradient in such disks leads 
to an instability, named the streaming instability. 
They suggested that the instability can produce 
local particle overdensities, which may help the 
formation of planetesimals. The simulations 
by Youdin and Johansen (2007) confirmed the 
instability and its clumping effect. Johansen 
and Youdin (2007) showed that, in the saturation 
stage of the instability, the effect is most prominent for 
marginally coupled particles with friction timescales 
close to the rotation period of the disk.  

Johansen et al.\ (2007) showed that including 
the particle feedback amplifies the maximum 
concentration from particle clustering 
by the MRI-driven turbulence in circumstellar disks. 
It suggests that the streaming instability 
from two-way coupling gives enhanced 
clustering strength in such disks. This appears 
to be different from the case of isotropic turbulence 
where the particle feedback reduces the high tails of the
concentration PDF. (We note, however,  
that the maximum particle density does not 
exceed 100 times the flow density in the 
simulations of Johansen et al.\ (2007) with no self-gravity).    
The amplification in the clustering intensity 
by the streaming instability was important for 
the planetesimal formation model of Johansen 
et al.\ (2007). A more detailed discussion of their 
model will be given in \S 6.3.  

\section{Clustering Statistics of Particles of Different Sizes}

So far we have only studied clustering of particles 
of the same size. In this section, we consider 
the relative spatial distribution of different 
particles. As mentioned earlier, the particle clustering 
location shifts in space as the particle size changes. 
This has interesting consequences for the clustering 
statistics of particles of different size. We quantify 
this effect by analyzing our simulation data. 
    
\subsection{The Bidisperse RDF}

We first compute the bidisperse RDF, $g(St_1, St_2, r)$, 
for two different particles with Stokes numbers 
$St_1$ and $St_2$, which is defined as the 
probability of finding a particle with $St_2$ (or $St_1$) at a 
distance $r$ from a reference particle with $St_1$ (or $St_2$). 
The computation is done in a similar way as for 
the mondisperse case. The RDF for the 
bidisperse case is equivalent to the two-point 
cross correlation function for the (number) 
density fields of two different particles.

\begin{figure}
\centerline{\includegraphics[width=1.0\columnwidth]{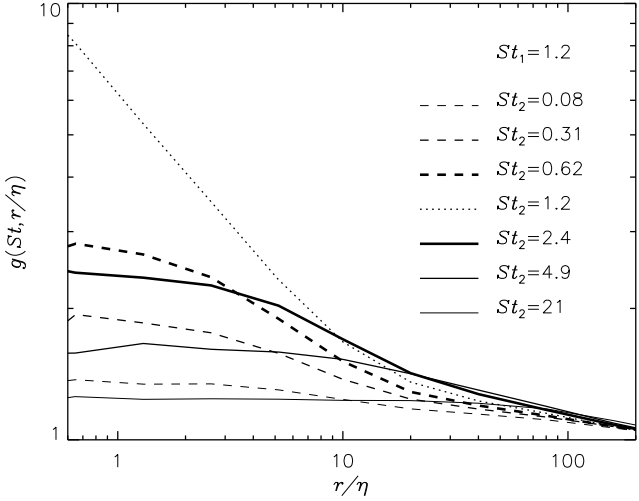}}
\caption{Bidisperse RDF for different Stokes number pairs. 
One of the Stokes numbers, $St_1$, is fixed at 1.2.  
The dotted line is the monodisperse  RDF for $St=1.2$. 
Thinner lines are used to plot RDFs with $St_2$ farther 
away from $St_1$.}
\vspace{0.1in}
\label{birdf}
\end{figure}

Fig.\ \ref{birdf} shows the bidisperse RDF as a 
function of the length scale for different Stokes 
number paris. One of the Stokes numbers is fixed 
at $St_1=1.2$, and the dotted line is the 
monodisperse RDF with $St=1.2$.  We see that, 
at large scales, the bidisperse RDF is close to the 
monodisperse one. This is because the particle 
clusters are generally located at the same regions 
when viewed at these large scales. With decreasing 
length scale, the bidisperse RDF becomes flat, 
consistent with results by Reade and Collins 
(2000b) and Zhou et al.\ (2001). This indicates 
the density fields of the two different particles 
become less correlated at smaller scales. 
The spatial separation between clustering 
locations becomes visible when examined 
at small scales. The length scale at which the 
RDF flattens increases as the ratio of the particle 
sizes increases, corresponding to a larger 
separation between the clustering positions of 
the two particles. Similar behaviors have been found for the 
bidisperse RDFs with other values for 
the fixed Stokes number $St_1$.  Chun et al.\ (2005) 
showed that the flattening trend exists as long 
as there is a difference in the particle sizes. Even 
if the Stokes numbers difference is small, one still 
finds a flat part of the RDF when going to sufficiently 
small scales, due to the finite (but small) shift in 
the clustering locations. This result suggests 
that, in the bidisperse case, the clustering effect 
contributes less to the particle collision 
rates than in the monodisperse case.  

\begin{figure}
\centerline{\includegraphics[width=1.0\columnwidth]{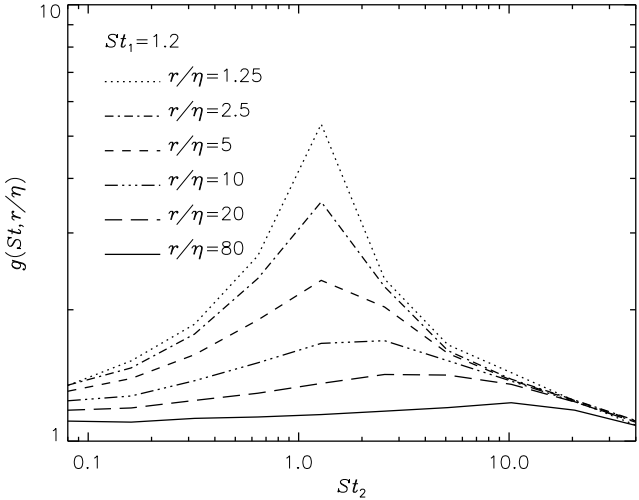}}
\caption{Bidisperse RDF at six length scales. $St_1$ is fixed at $1.2$.}
\vspace{0.1in}
\label{birdfscales}
\end{figure}

In Fig.\ \ref{birdfscales}, we show the bidisperse 
RDF as a function of $St_2$. The other Stokes 
number $St_1$ is fixed at 1.2. Different curves 
correspond to different length scales. For $\tilde{r} \sim1$, the bidisperse 
RDF peaks at $St_2 \sim St_1$, and decreases 
rapidly as the Stokes number ratio increases. 
The RDF is significantly reduced as the ratio 
increases to 3, and the density fields are 
essentially uncorrelated when the Stokes 
number ratio is larger than 10.  At larger length 
scales, the RDF peak moves 
to the right. This is because at these scales 
($r \gg \eta$) large particles ($St >1$) have 
stronger density fluctuations than the smaller 
ones ($St <1$).    

In summary, we found the bidisperse RDF 
becomes flat at small scales because particles of 
different sizes tend to cluster at different places. 
The bidisperse RDF decreases with increasing 
particle size ratio, and the effect of clustering on 
the particle collision rates between different 
particles is weaker than in the mondisperse case. 
The overall fluctuation amplitude of the particle density may 
be significantly suppressed if the particle 
size spanned an extended range.

\subsection{The Concentration PDF for Multiple Particle Sizes}

\begin{figure}
\centerline{\includegraphics[width=1.0\columnwidth]{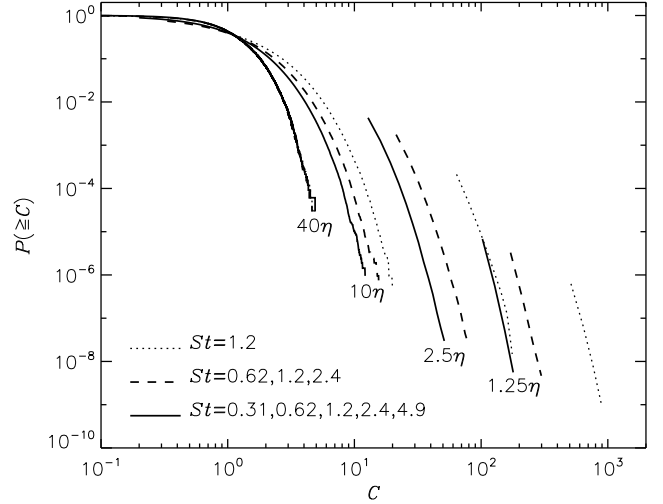}}
\caption{The cumulative concentration PDF at different length scales 
for multiple particle sizes. The dotted line corresponds to 
the monodisperse case with $St=1.2$, and the dashed and 
solid lines are for three and five particle sizes, respectively. }
\label{polypdf}
\end{figure}

In Fig.\ \ref{polypdf}, we show the concentration 
PDFs for two combinations of different particles 
with $St$ centered around $1.2$. 
For comparison, we also plot the PDF for the 
monondisperse case with $St=1.2$ (the dotted line). 
The dashed lines and the solid lines 
correspond to the results for combinations 
of 3 and 5 different particle sizes, respectively. 
The concentration factor shown in Fig.\ \ref{polypdf} 
represents the enhancement in the total {\it number} 
density in local regions relative to the 
average. When computing $C$ in each local 
region, we obtained the local number density by 
counting the total number of particles 
with size in the chosen range and 
then divided it by the average. 
Each particle size was given the same weighting 
factor as we have the same number of particles 
for each size in our simulations. The concentration 
$C$ computed this way can be understood 
as the enhancement factor in the particle 
mass density if the particle size distribution 
is such that the total mass of particles of each size 
is the same. Fig.\  \ref{polypdf} is just an illustration 
of how the concentration PDF changes in the 
presence of multiple particle sizes. For practical 
applications, one needs to use the proper 
weighting factor for each size according to the 
actual size distribution.

At $r=1.25\eta$, the PDF moves toward significantly smaller 
$C$, as the particle size range increases. There are two reasons 
for the behavior. First, at scales $\sim \eta$, 
the degree of clustering for each individual 
size decreases as the Stokes number gets 
farther from $1.2$. Including particles with 
$St$ larger or smaller than 1 leads to 
weaker overall clustering. Second, 
the bidisperse RDFs in Fig.\ \ref{birdf} show 
that if the Stokes number ratio of two particles 
is larger than 2, their clustering locations do 
not overlap at length scales $r \sim \eta$. 
This means that different particle sizes chosen 
in Fig.\ \ref{polypdf} essentially occupy 
different places when we look at scales $\sim \eta$. This 
has the effect of smoothing out the fluctuations in the 
overall particle density distribution, giving narrower PDF tails. 
The maximum $C$ in the measured 
PDFs at $r=1.25 \eta$ and $2.5 \eta$ for the 5-particle case 
is smaller than the monodisperse case (dotted line) by a factor of 
$4- 5$. This confirms that the strongest clusters 
of the 5 particles are spatially separated, with 
the typical separation larger than $\sim \eta$. 
The shift of the PDF tail toward smaller 
$C$ implies that the probability of  finding particle clusters 
of extreme concentration level is greatly reduced if the particle 
size has an extended range. Therefore, 
using a single typical (or average) particle size 
to approximate a size distribution could 
significantly overestimate the clustering 
intensity.  

As the length scale increases, the shift of the 
PDF tail becomes smaller, and at $r=40 \eta$ 
the tail is essentially unchanged. One reason 
is that the clustering locations of these 
particles overlap when viewed at $r=40 \eta$ (see Fig.\ \ref{birdf} 
and the discussion in \S 5.1).  Also unlike the 
case of small scales ($r \lsim \eta$) where the 
clustering intensity peaks at $St \simeq 1$, at $r= 40 \eta$
the clustering strength for particles of each individual 
size keeps increasing  as $St$ increases from 0.31 to 4.9 
(see Fig.\ \ref{monordfscales}). Therefore, 
the clustering intensity of larger particles 
(i.e., $St=2.4$, $4.9$) and smaller particles 
(i.e., $St=0.31$, $0.62$) is, respectively, 
higher and lower than that from particles 
of average size ($St=1.2$), and their contributions 
to the overall clustering can compensate 
each other. This explains why the PDF at $40\eta$ is  almost 
unchanged.   

We also computed the concentration PDFs for 
combinations of particle sizes centered around 
$St =10$. In that case, we found that the PDF 
at $40\eta$ for a combination of 5 different particles with 
$St=2.4, 4.9, 10, 21$ and $43$ is significantly narrower 
than the monodisperse PDF for $St=10$.  
The effect of the existence of multiple particle sizes 
on the overall clustering intensity at a given scale 
depends on both the average particle size 
and the width of the size distribution. The effect can 
be understood by considering whether the clustering 
locations of these particles overlap and how much 
each individual size contributes to the overall clustering. 
 
To summarize, we find that the presence 
of an extended particle size range tends to 
reduce the overall clustering intensity, and 
a careful consideration for the particle size 
distribution is needed to obtain an accurate 
estimate for the overall clustering intensity.

\section{Application to Protoplanetary Disks}

Turbulent clustering of inertial particles has 
potential applications to dust particles in many 
astrophysical environments, such as the interstellar 
medium, protoplanetary disks, and the atmospheres 
of planets and dwarf stars.  As mentioned earlier, 
clustering of dust grains may significantly increase 
their collision rates for particles of similar sizes, 
and thus needs to be considered in coagulation 
models. Here we will consider clustering of 
dust particles in protoplanetary disks and, in 
particular, its role in models for planetesimal 
formation. Applications to different environments may 
require exploring other complexities. For example, 
a study of dust grain dynamics in interstellar clouds needs 
to account for the Lorenz force due to the 
electrical charges on the grain surface and the 
presence of magnetic fields in the clouds. 

Preferential clustering of inertial particles in turbulence 
has attracted attention from the community of 
planet formation because it may provide a 
possible solution to the long-standing problem of 
planetesimal formation. As mentioned in the Introduction, 
the classic planetesimal formation theory is challenged by
the self-generated turbulent stirring, and growth of 
dust particles to kilometer  size by collisional coagulation 
suffers from the meter-size barrier.  Two potential solutions 
to this problem have been recently proposed by Johansen 
et al.\ (2007) and by Cuzzi et al.\ (2008). 
The model model by Cuzzi et al.\ (2008) is directly based on 
turbulent preferential clustering. 
Particle clumping in the simulations  of Johansen et al.\ (2007) 
may also have contribution from turbulent clustering. 
Before discussing these models, we first consider 
the properties of turbulence in protoplanetary disks. 

\subsection{Turbulence in Protoplanetary Disks}

Following Cuzzi et al. (2001), we use the $\alpha$ prescription 
for turbulence in the disks, i.e., the turbulent viscosity, $\nu_{\rm t}$, 
is parametrized as $\nu_{\rm t} = \alpha C_{\rm s} H$. 
An $\alpha$ value in the range of $10^{-3}$ to $10^{-5}$ 
is consistent with observations (see discussions in Cuzzi et  al.\ 2001, 2008). 
The scale height, $H$, of the disk is given by $H \simeq C_{\rm s}/\Omega_{\rm K}$ with 
$\Omega_{\rm K}$ being the Keplerian rotation 
frequency. The turbulent viscosity can be estimated 
from the turbulent rms velocity, $U$, and the integral 
scale, $L$,  by $\nu_{\rm t} \simeq UL$. Assuming the 
turnover time,  $\sim L/U$, of the largest turbulent 
eddies is of the order of $ \sim \Omega_{\rm K}^{-1}$ 
(Cuzzi et  al.\ 2001), we have $U = \alpha^{1/2} C_{\rm s}$ and $L = \alpha^{1/2} H$.

We assume a standard minimum-mass solar nebula where the density and 
temperature profiles are given by 
$\rho_{\rm g} = 1.7 \times 10^{-9} (R/\rm{AU})^{-2.75}$ g cm$^{-3}$
and $T = 280 (R/\rm{AU})^{-0.5}$ K.  
With these scalings, we find $U= 10^3 \alpha_{-4}^{1/2} (R/\rm{AU})^{-0.25}$ cm/s,  
and $L = 5 \times 10^4 \alpha_{-4}^{1/2} (R/\rm{AU})^{1.25}$ km,  
where $\alpha_{-4} \equiv \alpha/10^{-4}$. 
To calculate the Kolmogorov scales, we need to estimate 
the molecular viscosity, $\nu$. Assuming  a cross section 
of $2.5 \times 10^{-15}$ cm$^2$  for hydrogen molecules, we 
have $\nu = 5 \times 10^4 (R/\rm{AU})^{2.5}$ cm$^2$/s. 
We then obtain $\eta = 5 \times 10^3 \alpha_{-4}^{-1/4} ({R}/{\rm{AU}})^{2.38}$ cm, 
and $\tau_\eta = 5 \times 10^2 \alpha_{-4}^{-1/2}({R}/{\rm{AU}})^{2.25}$ s.
The friction timescale is given by $\tau_{\rm p} = 6 \times 10^3 ({a_{\rm p}} /{\rm{cm}}) ({R}/{\rm{AU}})^{3}$ s,
where we assumed the density of the dust material is 
$\rho_{\rm d} =1$ g cm$^{-3}$ and used the formula, eq.\ (2), for the Epstein regime. 
Finally,  we find the Stokes number is 
$St = 12 ( {a_{\rm p}}/{\rm{cm}}) ({R}/{\rm{AU}})^{3/4}$.   
Therefore, at 1 AU, the particle size with most intense 
turbulent clustering ($St \simeq 1$) is $a_{\rm p}  \simeq 0.1\alpha_{-4}^{-1/2}$ cm.

In our discussions below on planetesimal formation 
models, we will take the radius 5 AU as an example.  
Using the formulas given above, we find 
$U = 7 \alpha_{-4}^{1/2} $ m s$^{-1}$ and  $L = 4  \times 10^5  \alpha_{-4}^{1/2} $ km at 
5 AU.  
The Kolmogorov length and time scales are, respectively, 
$\eta \simeq 2 \alpha_{-4}^{-1/4}$ km 
and $\tau_\eta \simeq 2 \times 10^4 \alpha_{-4}^{-1/2}$ s. 
The friction timescale is given by $\tau_{\rm p} = 
8 \times 10^5 ({a_{\rm p}} /{\rm{cm}})$ s, and we 
have the Stokes number  $St = 40 \alpha_{-4}^{1/2} (a_{\rm p}/{\rm cm})$.
Therefore, at 5 AU the particle size corresponding to 
$St =1$ is $a_{\rm p} =  0.025 \alpha_{-4}^{-1/2}$ cm. 

As summarized by Brauer, Dullemond, and Henning (2008) 
and others, theoretical disk models, millimeter-wavelength 
observations, as well as a careful reevaluation 
of the minimum mass solar nebular by Davis 
(2006) all find radial density profiles flatter than the conventional -3/2 power law.  Brauer, 
Dullemond, and Henning (2008) adopt 
$\rho_{\rm g} \propto R^{-0.8}$, and, with this 
flatter density profile, the gas density at 1~AU would be 
30 times smaller  given the total mass of the disk. 
This gives changes to some quantities of interest here, 
which can be seen by examining their density 
dependence. For example, the friction timescale $\tau_{\rm p}$ scales 
with $\rho_{\rm g}$ as $\rho_{\rm g}^{-1}$ in the Epstein regime. 
The Kolmogorov length and time scales go like $\rho_{\rm g}^{-3/4}$ 
and $\rho_{\rm g}^{-1/2}$, respectively. More interestingly, 
we have $St \propto \rho_{\rm g}^{-1/2}$. Therefore, if the gas 
density at 1 AU is 30 times smaller, then the particle size 
with most intense clustering at 1 AU will be reduced by 
a factor of $5.5$, relative to the case of a standard 
minimum-mass solar nebula.

The optimal particle size for turbulent clustering in 
planetary disks may correspond to the size of 
chondrules, depending on the specific disc 
physical parameters. Cuzzi et al. (2001) suggested 
that turbulent clustering can play an important role in collecting 
and sorting chondrules in primitive chondritic meteorites. 
Using simulations that include multiple particle 
sizes, they showed that the particle size distribution 
in dense clusters produced by turbulent clustering 
is in good agreement with the size distribution 
of chondrules found in chondritic meteorites. However,
size sorting alone is not sufficient to explain the formation 
of large bodies (planetesimals or asteroids) with a 
significant fraction of their mass in the form of 
chondrule inclusions. Some other mechanism 
responsible for the aggregation of chondrules
into larger bodies (and also an explanation for 
the origin of their thermal processing) must be
envisioned. Cuzzi et al. (2008) proposed that large 
self-gravitating clusters of chondrule size particles 
could contract into planetesimals, which we discuss next.

\subsection{The model by Cuzzi et al. (2008)}

Cuzzi et al.\ (2008; hereafter C08) outlined 
a mechanism for planetesimal formation, 
based on dense clumps of chondrule-size 
particles produced by turbulent clustering. 
C08 first found that, due to the gas pressure 
and the fact that the chondrule-size particles 
are quite tightly coupled to the gas flow, 
even the densest clumps (with a local 
particle-to-gas ratio $\sim100$) cannot undergo 
a direct gravitational collapse. Instead, the self-gravity 
only leads to a gradual and gentle sedimentation 
of particles toward their mutual center. 
A slowly contracting clump is subject to 
various disruption mechanisms. An 
examination of the ram pressure 
disruption by head winds from the gas flow 
gives a constraint on the clump size. 
For a clump with the maximum loading factor 
of 100, its size is required to be larger than 
$\sim 10^4$ km in order for self gravity to 
be able to stabilize it. These persistent 
clumps would form ``sandpile" planetesimals 
of 10-100 km, once the particle sedimentation 
toward the clump center is complete.   
 
Cuzzi et al.\ (2010) and Chambers (2010) further 
developed this idea and gave quantitative 
predictions for the planetesimal formation 
rate and the initial mass function of asteroids 
(Cuzzi et al.\ 2010). The key element in these 
studies is the prediction of the probability of 
finding clumps of sufficient size ($\gsim 10^4$ km) 
and intensity (with local  $\Phi \sim 100$). 
From the calculations in \S 6.1, at 5 AU the 
integral scale is $L \simeq 4 \times 10^5$ km 
and the Kolmogorov scale $\eta \sim 2$ km, 
and thus the critical clump size $10^4$ km is 
within the inertial range of the disk turbulence. 
Therefore the prediction of the probability 
requires understanding of turbulent clustering 
at inertial-range scales, which, however, 
has not been well explored. In their quantitative 
predictions for that probability, Chambers (2010) 
and Cuzzi et al.\ (2010) made use of the cascade 
mode for the joint PDF of the particle concentration 
and the flow enstrophy developed by 
Hogan and Cuzzi (2007).

Before discussing the prediction of the cascade model, 
we first have a look at Fig.\ 1 in C08, which was 
used to illustrate the existence of strong particle 
clusters. From this figure, we can obtain a rough 
estimate of the probability of finding strong clumps 
of size $10^4$ km. The figure shows the 
concentration PDFs of $St=1$ particles for the case 
without the particle back reaction, including 
both the PDFs measured from the low-$Re$ 
simulations and those extrapolated to the realistic $Re$ 
values using the multifractal model by Hogan et al.\ (1990). As 
mentioned earlier, the latter is much broader. The PDFs in Fig.\ 1 of C08 
represent the probability of finding clumps of size $2\eta \simeq 4$ km. 
To estimate the probability for clumps of size $10^4$ km, we need to 
increase the length scale by a factor of $\sim 3 \times10^3$. 
In \S 4.2, we showed that the PDF tails become 
narrower with increasing length scales. For example, 
as the length scale increases from 2.5 $\eta$ to 40 
$\eta$ in our Fig.\ \ref{monopdf}, the PDF tail moves to the left, 
and the concentration $C$ in the high tail decreases 
by a factor of $\sim 50$. This indicates a very sensitive 
dependence of the PDF tail on the length scale. Increasing 
the length scale by a factor of $3 \times10^3$ 
(from $2\eta$ to $10^4$ km) would push the extrapolated 
PDFs in Fig.\ 1 of C08 to the left by 2 or 3 orders of 
magnitude, resulting in a narrow PDF at 10$^4$ km. 
The concentration level at the high tail (with $P(>C) 
\simeq 10^{-6}$) would be reduced to around or below $C\sim100$.
  
We also note that the particle concentration 
PDF shown in Fig.\ 1 of C08 is mass-weighted.
To estimate the probability of finding particle 
clumps of a given size, we need to use the 
volume-weighted PDF. This means that 
another correction is needed to account 
for the difference between the volume- and 
mass- weighted PDFs. This correction also 
gives a significant reduction because the 
volume-weighted PDF is narrower than the 
mass-weighted one (see \S4.2).  

The discussion shows that dense clumps 
of size $10^4$ km are quite rare, and the 
probability of finding such a clump is 
much smaller than the direct impression one 
may have from Fig.\ 1 of C08. The small 
probability is due to the narrow scale 
range (between the turbulence outer scale and $10^4$ km) 
available for clustering to proceed.  

We next argue that the cascade model used 
in the quantitative calculations of Chambers 
(2010) and Cuzzi  et al.\ (2010) may considerably 
overestimate the probability of finding large 
and dense particle clumps for planetesimal 
formation. In Appendix A, C08 preformed a 
24-level cascade for $St=1$ particles, and found 
a significant probability ($10^{-5} -10^{-6}$)  
for the existence of clumps with $C=1000$ 
(see Fig.\ 5 in C08). A 24-level cascade 
corresponds to a scale range of 256. 
Therefore, if the turbulent outer 
scale $L=4\times10^5 \alpha_{-4}^{1/2}$ km, 
the prediction was for the scale 
$\sim 2000 \alpha_{-4}^{1/2}$ km. 
We can roughly estimate the PDF tails at this 
scale by making adjustments, i.e., a length 
scale increase and the mass- to volume-weighting 
correction, to the tail of the extrapolated PDF 
at $2 \eta$ in Fig.\ 1 of C08 (see discussions above). 
It appears that the probability for $C=1000$ estimated 
this way is much smaller than predicted by the 
cascade model, suggesting a significant overestimate 
may exist in the cascade model prediction.     

We give a physical argument on why 
the cascade model may considerably 
overestimate the clustering intensity. 
The prediction of the cascade model 
depends on the multiplier PDF that 
controls each cascade step (\S 4.5). 
The multiplier PDF used in C08 and 
the followup studies was measured 
from a cascade step in the dissipation 
range, from $3 \eta$ to $1.5 \eta$ 
(see Hogan and Cuzzi 2007). This 
multiplier PDF was assumed to apply 
to all cascade steps including the 
steps in the inertial range.  

The validity of using the multiplier PDF 
from the $3 \eta$  - $1.5 \eta$ step in all 
cascade steps relies on the scale invariance of the 
multiplier PDF. Hogan and Cuzzi (2007) 
were concerned with this issue and 
gave some indirect evidence for this scale
independence. However, it was not 
directly verified, due to the limitations in 
the numerical resolution and the number 
of particles. In fact, it is reasonable to 
suspect the multiplier PDF may have a scale 
dependence considering that the density 
structures in the inertial range are not 
self-similar in simulations neglecting the 
back reaction. This non-similarity was 
seen from the RDF (\S 4.1) and the 
scale dependence of the singularity 
spectrum (see \S 4.3.2 and Cuzzi et al.\ 2001). 
As mentioned earlier, these results suggest 
that the clustering process occurs faster 
and faster as the length scale decreases 
toward $\eta$. Therefore, if one measures 
the multiplier PDF for the particle concentration 
in the simulations neglecting the back-reaction, 
its width would decrease with increasing length 
scales. If the cascade process has the same 
trend when the back-reaction is included, then the 
multiplier PDF in the inertial range would be narrower 
than that from the $3 \eta$ to $1.5 \eta$ step. This is of 
special concern for scales ($ \sim 10^4 \eta$) well 
separated from the Kolmogorov scale, $\eta$. 
Considering the large number of cascade steps, 
a slight overestimate in the multiplier PDF 
could result in a significant overestimate for 
the tail of the concentration PDF. 
There is a possibility that the inclusion 
of back-reaction may lead to a scale-independent 
multiplier PDF, but this remains to be verified.
  
The argument above suggests that the planetesimal 
formation rates calculated by Chambers (2010) and 
Cuzzi et al.\ (2010) using the cascade model could 
have been overestimated substantially. Chambers (2010) 
and Cuzzi et al.\ (2010) found that, to satisfy various 
constraints, the mean dust-to-gas ratio is required 
to be much larger than the standard value. If the cascade 
model overestimates the clustering intensity, then 
the required dust-to-gas ratio is even higher.    

C08 and the followup studies only considered a 
single particle size corresponding to $St =1$.    
As discussed in \S 5.2, the concentration PDF 
tends to become narrower if the particles size 
has a broader distribution around the average 
size.  Therefore, using a typical particle size to approximate 
the entire size distribution may significantly overestimate the 
probability of finding strong clusters. It is likely that 
dust particles in protoplanetary disks have an extended size 
range as a result of coagulation (see, e.g., Dullemond and Dominik 
2005). An accurate estimate for the probability 
requires a careful consideration of the effect of the 
particle size distribution on the clustering intensity.  

In \S 4.1, we found that, at a length scale $r$ 
in the inertial range, $St >1$ particles can 
have higher clustering intensity than $St=1$ 
particles. For $r \sim 10^4 \eta$, the particles 
that have strongest clustering would 
be those with $St \sim 300$, assuming the 
clustering length scale for $St >1$ particles 
is given by $l_{\tau_{\rm p}} \propto \tau_{\rm p}^{3/2}$. 
This Stokes number corresponds to a particle 
size of a few to ten cm at 5 AU. At the scale 
$10^4$ km, these particles would have stronger 
clustering than $St=1$ particles. Therefore, if the clustering 
intensity is the primary concern, particles with 
$St \sim 300$ may be a better candidate. 
However, the PDF of these particles at the 
scale of 10$^4$ km is also likely to be 
quite narrow.


In summary, the estimate of the probability of finding 
large and dense clumps needed to seed planetesimal 
formation requires an understanding of turbulent 
clustering at inertial-range scales. Our discussion 
suggests that the cascade model of Hogan and 
Cuzzi (2007) may significantly overestimate the 
probability of finding the required clumps, and 
thus Chambers (2010) and Cuzzi et al.\ (2010) 
may have overestimated the planetesimal formation rates by 
the C08 mechanism.  A future systematic study of the 
inertial-range clustering, accounting for various effects 
such as the presence of multiple particle sizes, is necessary to 
clarify the probability of finding particle clumps satisfying the 
constraints for planetesimal formation by the C08 mechanism.

\subsection{The model by Johansen et al.\ (2007)} 

Johansen et al.\ (2007, hereafter J07) 
carried out numerical simulations 
evolving meter-size boulders in the MRI-driven 
turbulence in protoplanetary disks. 
Dense particle clumps are observed 
in the simulations. In their runs neglecting 
the particle back reaction, the solids-to-gas 
ratio reaches a maximum of several tens in 
the densest particle clumps. 
The maximum concentration level is further 
amplified by an order of magnitude when 
the particle back reaction is included. 
The particle density in the densest clumps 
is  $\sim$100 times the local gas density, and these 
clumps of meter-size particles can undergo 
gravitational collapse, leading to rapid formation 
of planetesimals. In this subsection, 
we discuss various clustering mechanisms 
that contribute to the formation of particle clumps 
in J07 simulations. 

For realistic turbulent flows in rotating disks, 
three distinct scale ranges are expected. 
The first range is the large scales dominated 
by the rotation effects. The second one is the 
intermediate length scales regulated primarily 
by non-linear interactions, where the flow 
statistics are expected to be isotropic. 
The planetesimal formation model by 
Cuzzi et al.\ (2008) discussed in \S 6.2 
is based on turbulent clustering in this range. 
The last is the dissipation range at the smallest 
scales. The limited resolution in J07 simulations 
does not resolve scales in the intermediate 
range, and thus the rotation-dominated 
scales connect directly to a dissipation 
range corresponding to the hyper-diffusion term used in 
their simulations.  We discuss the clustering 
effects in these two scale ranges separately.  

In the rotation-dominated range, the effect 
of the Coriolis force is of particular interest. 
The Coriolis force has the effect of pushing 
particles toward the cores of anticyclonic 
vortices (whose vorticity is opposite 
to the disk rotation), leading to particle 
trapping in these vortices (e.g., Barge \& Somania 1995).  
Numerical studies in 2D found long-lived 
anticyclonic vortices, and particle trapping  
in these vortices was proposed to be a 
candidate mechanism for planetesimal 
formation (e.g., Bracco et al.\ 1999). 
However, for realistic 3D flows, the 
origin, the stability, and the lifetime of anti-cyclonic 
vortices have been debated (e.g., Johansen et al. 2004, 
Fromang \& Nelson 2005, Barranco and Marcus 2005 ).   

Numerical simulations have shown that 
long-lived large-scale zonal flows 
exist in MRI-driven turbulence (e.g., Johansen 
et al.\ 2009b, Fromang \& Stone 2009).  
Associated with these zonal flows are 
large-scale pressure bumps (where the vorticity is 
supposedly anticyclonic). The particle trapping 
capability of these pressure bumps was emphasized by 
Johansen et al.\ (2011). Comparing results from 
256$^3$ and $512^3$ simulations, 
they found numerical convergence for the 
strength of the pressure bumps and the particle 
trapping effect (Fig.\ 5 of Johansen et al.\ (2011)). 
However, the clustering intensity at 
large sales seems to be low.  

Fig.\ 2 of J07 shows the the formation 
process of gravitationally unstable clumps. More 
impressive than the large-scale clumps, 
strong particle clusters are seen at the 
smallest scales in the central four 
panels showing the particle concentration 
field before self-gravity is turned on. 
Some degree of radial contraction 
occurs around the small clumps after the 
gravity is turned on, and in a few rotation 
periods planetesimals form 
out of the contracted clusters. 
This suggests that the small-scale 
clusters may provide the primary seeds 
for the gravitational collapse of 
planetesimals, and thus clustering at small scales 
appears to be more important for the J07 model than 
the large-scale clumping. The maximum concentration 
factor reported in J07 is measured at the smallest scale 
in the simulations, corresponding to $\sim 1.6 \times 10^5$ km. 
A dense cluster at this scale can have sufficient 
mass to form a planetesimal of a fairly large size.

Given the apparent importance of small-scale 
clustering in J07, we now briefly discuss the 
clustering physics at small scales in the 
dissipation range of their simulations.
An important quantity for small-scale 
clustering is the Kolmogorov timescale. 
Using the rms vorticity provided 
by Johansen (2007, private communication), 
we calculated the Kolmogorov timescale, 
$\tau_{\eta}=\langle w^2 \rangle^{-1/2}$, 
which is 0.18 $\Omega_{\rm K}^{-1}$ in 
the 256$^3$ run. With this value of $\tau_\eta$,  the 
friction timescales ($(0.25 -1) \Omega_{\rm K}^{-1}$)
of the 4 particle sizes chosen in J07 
correspond to a Stokes number range 
$St \in (1.4-5.6)$. These numbers are 
close to unity, and thus turbulent clustering, in the 
sense of particle accumulation in strain-dominated regions (\S 2.1), 
may have considerable contribution to  the small-scale 
clustering in J07.    

Because the Kolmogorov timescale $\tau_\eta$ 
is only moderately smaller than the rotation 
period, the effect of the Coriolis force is not 
negligible even at the smallest scales in 
the J07 simulations, and it may also 
contribute to the small-scale clustering. 
We analyze the Coriolis effect from an derivation 
of the particle flow divergence using the 
same approach as in \S 2.1. The derivation 
yields two interesting terms. The first one 
is $\tau_{\rm p} (\omega^2/2 -s_{ij}s_{ij})$, 
which is the same as that given in \S 2.1 and thus 
corresponds to turbulent clustering. 
The second term is from the Coriolis 
force, and it is given by $2 \tau_{\rm p} \Omega_{\rm K} 
\omega_z$ where $\omega_z$ is the vorticity 
component in the direction of the disk rotation. 
This term reflects the trapping effect of the 
anticyclonic vortices. The derivation here 
is under the assumption that $\tau_{\rm p} < \tau_\eta$. 
This condition is not strictly satisfied for particles 
in J07. However, those particles have $St$ 
close to 1, and the derived divergence terms 
give a useful illustration for the Coriolis effect 
on particle clustering at small scales.   

We first point out that the Coriolis term only 
acts on vorticity in the $z$-direction, and it does 
not affect turbulent clustering due to vortices in other directions. 
For vorticity in the vertical direction, the amplitude 
of the Coriolis term, $2 \tau_{\rm p} \Omega_{\rm K} \omega_z$, 
is close to the vorticity term, $\tau_{\rm p} \omega^2/2$, 
for turbulent clustering, since $\omega \simeq 5 \Omega_{\rm K}$. 
This suggests that, in anticylonic vortices, the Coriolis 
force is strong enough to resist the particle expelling effect 
from turbulent clustering. On the other hand, 
the Coriolis force helps turbulent clustering 
push particles out of the cyclonic vortices. 
The opposite effects of the Coriolis force in 
anticyclonic and cyclonic vortices may 
cancel each other, and the clustering intensity 
in the strain-dominated regions may be similar to 
that due to the turbulent clustering effect alone. 

The contribution from turbulent clustering 
is artificial in the sense that, due to the 
limited resolution, the friction timescale 
of the chosen particles happens to be close the 
smallest timescale in the simulations. 
As discussed in \S 4.4,  for particles with 
an artificial $St$ value close to 1 in a 
simulated flow, the clustering intensity is likely to 
decrease as the numerical resolution (or $Re$) increases. 
This is because $St$ becomes larger with 
increasing $Re$, causing a reduction 
in the clustering strength (see \S 4.4 for
a detailed discussion). In the real flow, 
the meter-szie particles chosen by J07 
have huge Stokes numbers, $St \simeq 2000$ at 5 AU, and for 
such large $St$ the contribution from turbulent clustering is likely 
to be negligible. As for the effect of the Coriolis force on the small-scale clustering,
it is not clear how it would change with increasing 
resolution.

Johansen et al.\ (2011) conducted simulation 
runs at two resolutions, 256$^3$ and 512$^3$. 
Their Fig.\ 7 shows that the total mass 
of gravitationally bound clumps and the mass of 
most massive clumps as a function of time in the two runs. 
Although it converges at late times, the total mass is 
smaller in the 512$^3$ simulation than 
in the 256$^3$ one in the early stage, 
when the formation of bound objects 
primarily depends on dense particle 
clusters. The mass of the most massive clumps 
in the 512$^3$ run is also smaller 
than in the 256$^3$ run at all times. 
This could be due to the numerical
reasons given in Johansen et al.\ (2010). 
The other possibility is that the small-scale 
clustering is actually less strong in the $512^3$ run 
where the Stokes numbers are larger. 
The latter would be expected if turbulent 
clustering were the dominant (though artificial) 
clustering mechanism at small scales.  

We finally discuss the effect of particle back-reaction 
on the clustering intensity. As mentioned earlier, J07 found 
that including the back-reation significantly 
increases the clustering amplitude. 
This was argued to be due to the streaming instability 
(Youdin and Goodman 2005; see discussions in \S 4.5). 
The particles chosen in J07 are marginally 
coupled to the disk rotation, and the effect 
of the streaming instability was shown 
to be most prominent for these particles 
(Johansen and Youdin 2007). Bai and 
Stone (2010) studied numerical 
convergence for the steaming instability 
with 2D simulations neglecting vertical 
stratification, and found that the particle 
concentration PDF converges at the 
resolution of $2048^2$. However, due to the 
peculiar properties of 2D turbulent flows, it is not clear 
if a similar convergence would also be found in 3D. 
If such a numerical convergence is confirmed by 
future 3D simulations including the effect of 
vertical stratification, then, as the numerical 
resolution increases,  the streaming instability 
can maintain sufficiently high clustering intensity for 
the planetesimal formation mechanism by J07, even though the contribution 
from turbulent clustering at small scales would decrease. 
      
To summarize, we argued that, along with 
other clumping mechanisms, turbulent 
clustering gives considerable contribution 
to the small-scale clumps in the 
simulations by J07. This contribution is 
due to the limitation in the numerical 
resolution, and would probably decrease 
as the resolution increases. Future work 
is needed to investigate how the small-scale 
clustering intensity changes with 
numerical resolution, and hence quantify the 
relative importance of turbulent clustering.

\section{Conclusions}

We have studied the spatial clustering of inertial 
particles suspended in turbulent flows using 
numerical simulations. We have presented a detailed 
analysis of the clustering statistics for 
11 particle sizes covering the approximate 
Stokes number range $0.1 \lsim St \lsim 100$. 
From the simulation data, we have measured 
the radial distribution function and the probability 
distribution function of the particle 
concentration. Our main results are 
summarized as follows. 

\begin{enumerate}

\item
For $St \lsim 1$, the clustering intensity increases with $St$, 
and very strong clustering is found in the dissipation range. 
On the other hand, if $St > 1$ and $\tau_{\rm p}$ 
corresponds to an inertial-range timescale in the 
turbulent flow, clustering occurs primarily at an 
inertial-range scale, $l_{\tau_{\rm p}}$. 
The clustering intensity at  the scale $l_{\tau_{\rm p}}$ 
decreases with increasing $St$. At scales below 
$\sim \eta$, the RDF has a strong peak 
at $St \simeq 1$ and decreases rapidly 
as $St$ gets away from $1$. At a given 
inertial-range scale, the maximum clustering 
intensity is from particles with $\tau_{\rm p}$ 
in the inertial range.  

\item
For $St \sim 1$, the RDF increases rapidly 
toward smaller scales and reaches large 
values at scales well below the 
Kolmogorov scale, $\eta$. This suggests that 
turbulent clustering can strongly increase the particle 
collision rates due to the enhanced probability 
of finding nearby particles. At all Stokes 
numbers, the RDFs below $\eta$ follow power-laws, 
and the scaling exponent, $\mu$, peaks at $St \simeq 1$. 
The increase of the RDF can continue to the larger 
of the two scales: the Brownian scale or the particle size.   
  
\item  
At small scales ($\sim\eta$), particles with $St \simeq 1$ 
show the broadest PDF tails. In our 512$^3$ simulation 
with $Re_{\lambda} \simeq 300$, the PDF tail of $St =1.2$ 
particles reaches $C \sim 100-1000$ at $r \sim \eta$. 
The PDF width for these particles decreases rapidly 
as the length scale increases, consistent with the RDF results.  
At inertial-range scales, the PDF width peaks at a Stokes 
number in the inertial range (i.e., $St>1$), and the Stoke 
number with maximum PDF width increases with 
increasing length scale. This suggests that, at length scales 
relevant for  the formation of planetesimals in 
protoplanetary disks, the strongest clustering would be 
achieved by particles with $St$ much larger than 1. 

\item
Consistent with previous studies, the bidisperse 
RDF between particles of different sizes becomes 
flat at small scales because different particles 
tend to cluster at different locations. The contribution 
from turbulent clustering to the collision rates 
between different particles is weaker than that 
between identical particles. The spatial drift of 
the clustering location as the particle size changes 
has the effect of smoothing the overall spatial 
distribution of the particles. This tends to make the particle 
concentration PDF narrower if the particle size distribution 
spans an extended range. Using a typical size instead 
of the actual size distribution may significantly 
overestimate the overall clustering intensity. 

\end{enumerate}

Several recent studies have proposed that strong  
clustering of dust particles in protoplanetary 
disks could provide a solution to the 
problem of planetesimal formation. The model 
of Cuzzi et al.\ (2008) is based on particle clusters of 
sufficient size ($10^4$ km) and concentration level (with local mass loading factor 
$\Phi_{\rm m} \sim$100) produced by turbulent clustering. 
The probability of finding these strong clusters 
depends on the clustering statistics at inertial-range 
scales, which are not well understood. 
We pointed out that the cascade model used by 
Cuzzi et al.\ (2010) and Chambers (2010) may 
significantly overestimate this probability and 
hence the predicted planetesimal formation 
rate. Further numerical studies are needed to 
better quantify the amplitude of turbulent 
clustering in the inertial range and set firmer 
constraints on this planetesimal formation model.

We discussed various clustering mechanisms
in the planetesimal formation simulations of 
Johansen et al.\ (2007, 2010). We argued 
that turbulent clustering may have considerable 
contribution in these simulations, because 
the particle sizes chosen in the study happen 
to have $St$ around unity due to the limited 
numerical resolution. The contribution is likely 
to decrease with increasing resolution and 
become negligible as the Reynolds number 
increases to its realistic value. Further 
numerical work should establish that particle 
clustering by other mechanisms in the simulations 
by Johansen et al.\ (2007), such as the particle 
trapping effect by the Coriolis force and the 
streaming instability, remains intense as the 
resolution increases, allowing the formation of 
planetesimals despite the reduced effect of 
turbulent clustering.
  
While our study provides a detailed analysis of 
the statistics of turbulent clustering, several 
important questions remain to be answered 
by future work. Our discussion on the Reynolds 
number dependence of the clustering properties 
was based on a review of previous numerical 
studies. A definite result on the Reynolds 
number dependence requires simulations with 
higher numerical resolution. A larger number 
of particles would reduce the Poisson noise 
and help improve the accuracy of the statistical 
measurements, especially for less clustered 
particles. Clustering statistics at inertial 
range scales are of special interest to 
planetestimal formation models based on turbulent clustering, 
and deserve a careful and thorough exploration. 
We neglected back reaction from the particles to 
the carrier flow in our simulations. A systematic 
study of the back reaction effect on both the 
RDF and the PDF for particles over an 
extended Stokes number range would help 
better understand the role of turbulent clustering in 
various astrophysical environments.   

\acknowledgements

We thank the referee, Anders Johansen, for an 
exhaustive referee report that helped us improve 
the paper. LP thanks Evan Scannapieco for helpful 
comments and acknowledges support by the 
NASA theory grant NNX09AD106. PP is supported 
by MICINN (Spanish Ministry for Science and Innovation) 
grant AYA2010-16833 and by the FP7-PEOPLE-2010-RG
grant PIRG07-GA-2010-261359. JS acknowledges support 
by the NASA Astrobiology Institute, Virtual Planetary 
Laboratory Lead Team. AK is supported in part by the 
National Science Foundation under grant AST0908740. 
The simulation utilized NSF TeraGrid resources provided by SDSC through 
allocation MCA07S014. 

\appendix

\section{A. Particle Clustering in Burgers Vortex}

\begin{figure}
\centerline{\includegraphics[width=0.55\columnwidth]{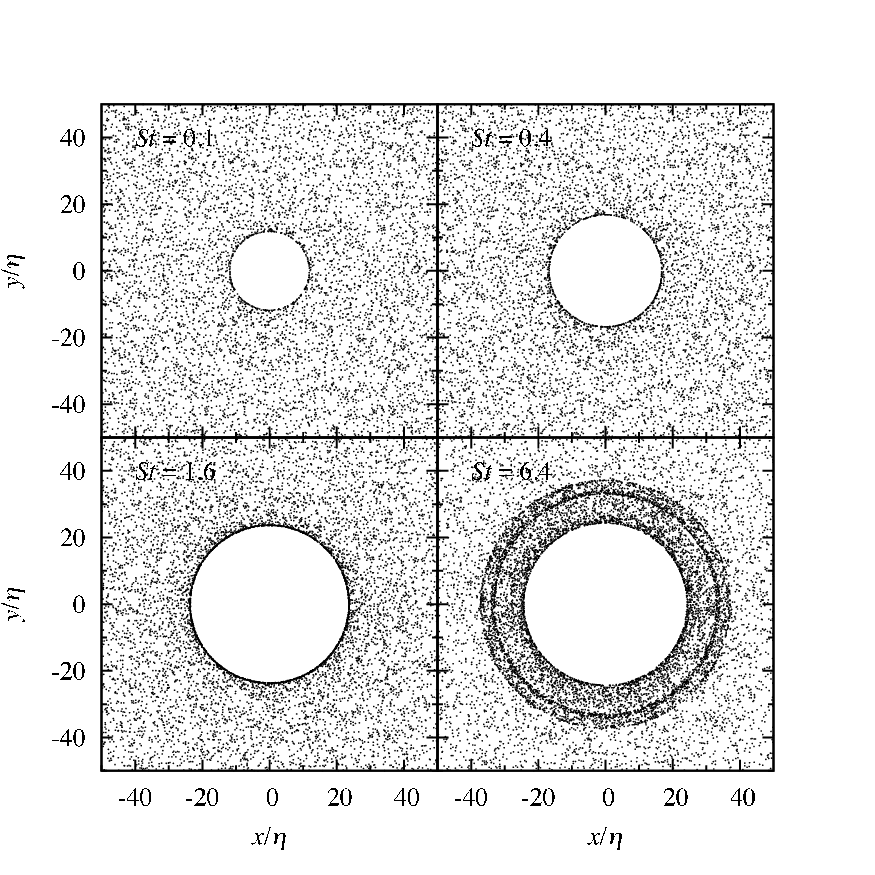}}
\caption{Spatial distribution of inertial particles in a Burgers 
vortex tube with $r_0 = 6 \eta$ and 
$U_0 = 14 u_\eta$.
The four panels correspond to particles of different 
Stokes numbers ranging from 0.1 to 6.4.}
\label{burgers}
\end{figure}

The physics of turbulent clustering of inertial particles  
described in \S 2 can be illustrated by 
a simple example using Burgers vortex tube as a model for 
the small-scale velocity structures in turbulent flows. 
This example provides insight into the relative spatial 
distribution between particles of different sizes.    

Vortex tubes are found to be fundamental 
building blocks in incompressible turbulence flows. 
Visualizations of the vorticity field 
in high-resolution simulations show 
numerous tube-like vortex structures. 
We use Burgers vortex, an exact solution of 
the Navier-Stokes equation, as a model for these tubes. 
The velocity in a Burgers vortex is given by,  
\begin{equation}
\begin{array}{lll}
{\displaystyle u_r =  -A r}\\
{\displaystyle u_{\theta} = \frac{\Gamma}{2\pi r} \left(1-exp\left(-\frac{Ar^2}{2 \nu}\right) \right)}\\
{\displaystyle u_z = 2A z}
\end{array}
\end{equation}
where $r$ is the radial distance to the tube 
axis, $\nu$ is the kinematic viscosity, $A$ is the 
strain that drives the vortex, and $\Gamma$ 
is the circulation of the vortex. The circulation 
velocity, $u_{\theta}$, has a maximum, $U_0$, 
at a radius $r_0 = 1.585 (\nu/A)^{1/2}$. 
This radius and the maximum circulation 
velocity have been measured by 
experiments, and the two parameters, 
$A$ and $\Gamma$, can be converted from $r_0$ and $U_0$. 
In our illustrative example, we adopt  $r_0 = 6 \eta$ 
and $U_0 = 14 u_\eta$ from the  
experimental results by Mouri et al.\ (2007) 
for intense tubes in a turbulent flow with Taylor Reynolds 
number $Re_{\lambda} \simeq 2000$ 
($Re \simeq 10^5$).   
We also tried different values for the parameters 
and found qualitatively similar behaviors for 
the particle spatial distribution.

The motion of an inertial particle in a Burgers 
vortex is determined by the competition 
between the drag toward the tube 
center by the radial flow ($u_r$) and 
the centrifugal force from the rotation 
induced by the circulation velocity ($u_\theta$). 
For a particle released at a 
large distance from the tube axis, 
the radial drag dominates at first.  
As it moves closer to the center, the 
particle rotates faster. When the 
centrifugal force from the rotation 
balances the radial drag, the particle ends up in 
a steady-state orbit (Marcu et al.\ 1995). Very small 
particles may not have steady-state orbits, 
instead they reach the tube center 
because of the efficient radial drag. The 
steady-state radius, which we will 
refer to as the equilibrium radius, 
can be estimated by $u_\theta^2/r = A r/\tau_{\rm p}$ 
(Marcu et al.\ 1995). Using $u_\theta$ as a 
function of $r$ in eq.\ (A1), we see that larger 
friction timescales give larger equilibrium radii. 
The particle motion in the $z$-direction 
is decoupled from that in the 
$r$-$\theta$ (or $x$-$y$) plane.  

In Fig.\ \ref{burgers}, we show the particle distribution 
in a vortex tube.  The particles are 
released at a constant rate from a 
cylinder at a distance of 100 $\eta$ from 
the center. At the time of release, we set the particle 
velocity to be the same as the flow velocity. 
For small particles, a ring forms at the 
equilibrium radius. Inside the ring, there 
are no particles because of the ejection by 
vorticity. 
With increasing friction timescale, the radius 
of the ring increases and more particles 
accumulate in the ring, leading to a larger particle 
density there. This results in a stronger clustering 
effect at larger Stokes number 
for $St \lsim 1$. For Stokes numbers much larger than 1, 
we find expanded ``rings" around 
the equilibrium radius.  A large particle can
overshoot the equilibrium radius because of 
its long memory of  the flow radial velocity. 
When it is dragged back by the 
radial flow from the other side, 
it overshoots the equilibrium radius again. 
This produces a ``ring" that is quite spread out, 
and as a consequence it has a smaller 
density than in the thin rings for $St \simeq 1$.  
For $St > 1$, the ring becomes thicker as $St$ increases, 
and the clustering intensity decreases. This simple 
example thus provides an intuitive explanation for 
why the maximum clustering occurs at 
$St \simeq 1$.

The example also offers insight into the clustering 
statistics for particles of different sizes. As seen 
from Fig.\ \ref{burgers} different particles have 
different equilibrium radii around the vortex tube. 
This suggests that clusters of different particles 
are located at different places in the flow. 
The density fluctuations of two different particles 
would be uncorrelated at scales below the typical 
separation between their clustering 
locations. The effect is especially strong for $St \lsim 1$ 
particles. The implication of this effect are discussed 
in details in the text.

\section{B. The Brownian Scale}

Collisions with flow molecules induce Brownian 
motions of the particles, which would diffusively spread 
particles in space, and limit clustering at 
small scales. In this Appendix, we estimate the 
Brownian scale, $l_{\rm B}$, below which clustering is 
suppressed by Brownian motions.  

The Brownian scale is essentially 
controlled by the competition of two effects: 
diffusive Brownian motions and the 
compressibility in the collective particle 
motions. We calculate the Brownian scale by 
estimating how far Brownian motions transport 
a particle during a timescale, $\tau_{\rm c}$, 
characteristic of the rate of compression/expansion 
in the particle flow. If the Brownian diffusion 
coefficient is $D$, we have $l_{\rm B} \simeq (D \tau_{\rm c})^{1/2}$.  

The diffusion coefficient $D$ can be derived 
from the Langevin equation (see, e.g., Gardiner 2004), 
$D = v_{\rm B}^2 \tau_{\rm p}$, where the 
Brownian speed, $v_{\rm B}$, is given by 
$(k T/m_{\rm p})^{1/2}$ assuming a thermal 
equilibrium between the gas molecules 
and the inertial particles. We estimate the 
characteristic compression timescale, 
$\tau_c$, by $(\partial_i v_i)^{-1}$. 

Using  eq.\ (4) for the particle flow divergence 
for $St<1$ gives $\tau_c \simeq \tau_\eta^2/\tau_{\rm p}$, 
and we have,  
\begin{equation}
l_{\rm B} = v_{\rm B} \tau_{\eta} {\rm ,}\hspace{5mm} 
{\rm for}\hspace{2mm} St<1. 
\label{lb}
\end{equation}
Note that $l_{\rm B}$ is actually the scale at 
which the flow velocity difference equals 
the Brownian speed. This is expected since 
no particle clustering would occur if the 
relative speed between particles is dominated 
by the contribution from Brownian motions. 
Typically, $l_{\rm B}$ is much smaller than $\eta$ 
because $v_{\rm B}$ is usually smaller than 
$v_\eta$. This allows strong clustering deep 
in the dissipative range.

Similarly, using the effective compressibility 
for $St>1$ particles given in \S 2.1, we obtain  
\begin{equation}
l_{\rm B} = v_{\rm B} \tau_{\rm p}{\rm ,}\hspace{5mm} {\rm for}\hspace{2mm} St > 1.
\label{lb2}
\end{equation}   
The effective compressibility for $St>1$ was evaluated using rough 
assumptions, and thus the Brownian 
scale given here should also be taken 
as a rough estimate. 
However, from Fig.\ \ref{monordf} we see that the RDF for $St \gsim 10$ is 
flat below the scale $l_{\tau_{\rm p}}$ and 
the degree of clustering does not change 
significantly toward small scales, and 
thus an accurate estimate of $l_{\rm B}$ is not crucial 
for $St \gg 1$.   

Balkovsky et al.\ (2001) suggested that particle clustering is suppressed 
below a diffusion scale defined as $l_{\rm D} = (D \tau_\eta)^{1/2}$.  
This scale is the same as that defined in the context 
of turbulent mixing of passive tracers. The Brownian scale 
we derived is different from the diffusion scale, $l_{\rm D}$, for $St \ne 1$. 
We argue the Brownian scale defined here is more 
appropriate for the suppression of particle clustering by Brownian 
motions.    

In the context of turbulent mixing, the diffusion scale is the smallest scale 
where scalar fluctuations exist. It reflects 
the competition between turbulent stretching, which 
produces structures at progressively smaller scales, and the 
molecular (or Brownian) diffusion, which tends to 
smooth out the structures. Here we are 
interested in the scale where the maximum 
clustering intensity is achieved. 
It represents the effect of Brownian 
motions on suppressing the growth of 
the fluctuation intensity by turbulent clustering. Although it can transfer the fluctuation 
power toward small scales,  turbulent stretching 
does not enhance the overall fluctuation amplitude. 
In particular, it does not give rise to a power-law increase 
in the clustering amplitude toward small scales. 
Therefore, it  does not play a role in determining the Brownian 
scale where the clustering amplitude reaches the maximum.      

Comparing $l_{\rm B}$ with $l_{\rm D}$, 
we find that $l_{\rm B} = l_{\rm D}/St^{1/2}$ for $St <1$ and $l_{\rm B} = St^{1/2} l_{\rm D}$ 
for $St>1$. Therefore $l_{\rm B}$ is larger than $l_{\rm D}$ at 
all $St$, except at $St =1$. This means that the particle density structures can 
exist below $l_{\rm B}$ due to turbulent stretching.  However, the power-law increase 
toward smaller scales would end at $l_{\rm B}$, below which 
the fluctuation amplitude does not significantly increase any more.

\end{document}